\documentclass[twocolumn,nofootinbib]{revtex4-1}
\usepackage[utf8]{inputenc}
\usepackage{amsmath}
\usepackage{graphicx}
\usepackage{amssymb}
\usepackage{color,colortbl}
\usepackage{cancel}
\usepackage{framed}
\usepackage{comment}
\usepackage[margin=0.7in]{geometry}
\usepackage{mathtools}
\usepackage{hyperref}
\usepackage{chngcntr}
\usepackage{soul}
\usepackage{subfig}

\counterwithin{equation}{section}

\makeatletter

    \def\CT@@do@color{%
      \global\let\CT@do@color\relax
            \@tempdima\wd\z@
            \advance\@tempdima\@tempdimb
            \advance\@tempdima\@tempdimc
    \advance\@tempdimb\tabcolsep
    \advance\@tempdimc\tabcolsep
    \advance\@tempdima2\tabcolsep
            \kern-\@tempdimb
            \leaders\vrule
                    \hskip\@tempdima\@plus  1fill
            \kern-\@tempdimc
            \hskip-\wd\z@ \@plus -1fill }
    \makeatother

\newcommand{\beq}{\begin{equation}}
\newcommand{\eeq}{\end{equation}}
\newcommand{\bea}{\begin{eqnarray}}
\newcommand{\eea}{\end{eqnarray}}

\begin{document}


\date{\today}

\title{\Large \bfseries Probing Bosonic Stars with Atomic Clocks}
{\raggedleft CERN-TH-2019-145 \\}
{\raggedleft CP$^\textnormal{3}$-Origins-2019-20 DNRF90 \\}
\author{Chris Kouvaris}
\email{kouvaris@cp3.sdu.dk}
\affiliation{CP$^3$-Origins, Centre for Cosmology and Particle Physics Phenomenology University of Southern Denmark, Campusvej 55, 5230 Odense M, Denmark}
	\affiliation{Theoretical Physics Department, CERN, 1211 Geneva, Switzerland}
\author{Eleftherios Papantonopoulos}
\email{lpapa@central.ntua.gr}
\affiliation{National Technical University of Athens, Zografou Campus 9, Heroon Polytechneiou Str. 157 80 Athens, Greece}
\author{Lauren Street}
\email{streetlg@mail.uc.edu}
\affiliation{University of Cincinnati, Dept. of Physics, Cincinnati, OH 45221 USA}
\author{L.C.R. Wijewardhana}
\email{rohana.wijewardhana@gmail.com}
\affiliation{University of Cincinnati, Dept. of Physics, Cincinnati, OH 45221 USA}

\begin{abstract}
Dark Matter could potentially manifest itself in the form of asymmetric dark stars. In this paper we entertain the possibility of probing such asymmetric bosonic dark matter stars by the use of atomic clocks. If the dark sector connects to the standard model sector via a Higgs or photon portal, the interior of boson stars that are in a Bose-Einstein condensate state can change the values of physical constants that control the timing of atomic clock devices. Dilute asymmetric dark matter boson stars passing through the Earth can induce frequency shifts that can be observed in separated Earth based atomic clocks.  This gives the opportunity to probe a class of dark matter candidates that for the moment cannot be detected with any different conventional method.
\end{abstract}

\maketitle

\section{Introduction}
Currently, cosmological and astrophysical observations leave little doubt about the presence of dark matter (DM) in our Universe~\cite{Aghanim:2018eyx,Bertone:2004pz}. Although its existence is very well motivated, the nature of DM still remains a complete mystery. The masses of possible DM candidates span several orders of magnitude ranging from ultralight particles of $\sim 10^{-22}$ eV~\cite{Hu:2000ke,Arvanitaki:2009fg,Hui:2016ltb}, to massive black holes of tens of solar masses \cite{Frampton:2009nx,Frampton:2010sw,Garcia-Bellido:2017fdg}. Furthermore, it is possible that DM  consists of several different components. Currently the so-called Collisionless Cold Dark Matter (CCDM) paradigm is consistent with observations of the large scale structure, suggesting that DM self-interactions are absent or very small. On the other hand, observations of the small scale structure seem to be at odds with CCDM. The core-cusp problem of dwarf galaxies, the diversity problem, and the ``too big to fail" raise doubt about the validity of the CCDM paradigm (see \cite{Tulin:2017ara} and references therein). Although these issues can be attributed to different factors, self-interacting DM (SIDM) can alleviate these problems, reconciling theory with small scale structure observations.  
Several studies of SIDM have been undertaken \cite{Vogelsberger:2012ku,Rocha:2012jg,Zavala:2012us,Peter:2012jh} providing constraints and an optimal range of cross sections, $0.1-10~\text{cm}^2/\text{g}$, for DM self-interactions that can solve the CCDM problems. Stringent constraints are imposed, for example, for the case where the self-interactions are mediated by a particle $\phi$ which is coupled to the Standard Model (SM) via  a Higgs portal \cite{Burgess:2000yq,Patt:2006fw,Andreas:2008xy,Andreas:2010dz,Djouadi:2011aa,Pospelov:2011yp}. In such a case, $\phi$ must decay before the start of the Big Bang Nucleosynthesis (BBN) in order to avoid energy injection to the plasma during BBN. These constraints can be evaded if $\phi$ is also coupled to sterile or active neutrinos \cite{Kouvaris:2014uoa}.  Also, SIDM might be needed in order to provide seeds for the existing supermassive black holes we observe in the Universe~\cite{Pollack:2014rja}. Finally, SIDM is motivated if one embeds the DM sector in Grand Unified Theories (GUTs) \cite{Kamada:2019gpp}.  The punchline is that SIDM might be welcome as it can alleviate problems in the CCDM paradigm and/or explain unresolved astrophysical issues.

One particular class of SIDM theories is that of asymmetric DM (ADM).  ADM is an alternative paradigm to thermally produced dark matter such as Weakly Interacting Massive Particles (WIMPs). In the usual WIMP paradigm, DM annihilates to SM particles. It turns out that for an annihilation cross section on the order of the weak interactions, DM annihilations reduce the DM relic density to the value observed today. This is the so-called WIMP miracle. However, this is not the only theoretically motivated production mechanism. Another interesting one is that of ADM. In this case, an asymmetry between the number of DM particles and antiparticles is created in the early Universe. Strong DM annihilations deplete the population of the particles in lack, leaving only DM particles of the species in excess. This is also a very well motivated paradigm. For example one can imagine a common asymmetry mechanism for baryogenesis and DM genesis. If for any baryon unit asymmetry, a DM unit  is also created, then a DM particle of a mass $\sim$5 GeV provides the correct relic abundance of DM in the Universe.  For a review on ADM see~\cite{Petraki:2013wwa}.

An ADM component that exhibits self-interactions among DM particles can cause the collapse of a DM cloud either by a gravothermal collapse mechanism such as in~\cite{Pollack:2014rja} or by effectively evacuating the energy of the system via  ``dark"-Bremsstrahlung radiation~\cite{Chang:2018bgx}, leading to black hole or asymmetric dark star formation. The latter are stable compact objects where their hydrodynamic stability is caused by DM self-interactions, Fermi pressure or the uncertainty principle depending on the underlying model and the nature of the DM particle (i.e. if it is  a fermion or a boson). The possibility of forming compact stable objects consisting of  fermionic ADM with self-interactions was studied in \cite{Kouvaris:2015rea}.
The Tolman-Oppenheimer-Volkoff equation (see \cite{Kouvaris:2015rea} and references therein) was solved and the mass-radius relation was found for these objects. It was assumed that the self-interactions were Yukawa-type and could either be attractive mediated by a scalar field $\phi$ or repulsive mediated by a vector boson field $\phi_{\mu}$.  The case of bosonic SIDM  forming compact star-like objects was studied in \cite{Eby:2015hsq}, where the density profile, the mass-radius relations, and the maximum mass that these objects can withstand were derived. We should stress that asymmetric dark stars are distinctly different from dark stars that might have formed in the past if DM is of symmetric nature~\cite{Spolyar:2007qv,Freese:2008hb,Freese:2008wh}. In the latter case, the hydrodynamic stability of the star is achieved by radiation pressure from the DM annihilation. These stars, if they ever existed, should have annihilated by now. On the contrary, due to the particle-antiparticle asymmetry, no annihilation takes place inside the asymmetric dark star. The species in excess have already annihilated away the minority component early on. Therefore, once formed asymmetric dark stars can be stable. We should also add that dark stars can exist in the form of hybrid compact stars made of baryonic and DM~\cite{Leung:2011zz,Leung:2013pra,Tolos:2015qra,Mukhopadhyay:2015xhs} as well as in the context of mirror DM~\cite{Khlopov:1989fj,Silagadze:1995tr,Foot:1999ex,Foot:2000iu}.  

In the case of ``bosonic" stars, if matter is sufficiently cold it stays in the ground state which is a Bose-Einstein condensate (BEC) state. 
Several boson particles can form bosonic stars, e.g., axions, or the scalars which drive the expansion of the Universe in quintessence models \cite{PhysRev.172.1331,PhysRev.187.1767,Breit:1983nr,Colpi:1986ye,Ratra:1987rm,PhysRevLett.103.111301,Barranco:2010ib,Schive:2014dra,Schive:2014hza,Eby:2014fya,Guth:2014hsa,Davidson:2016uok,PhysRevLett.117.121801,Chavanis:2016dab,Eby:2016cnq,Visinelli:2017ooc,Eby:2017teq,Eby:2018dat,Chavanis:2011zi,Chavanis:2011zm,Braaten:2018nag,Eby:2019ntd,Guerra:2019srj}.  Recently, the authors of \cite{Aharony:2019iad,Banerjee:2019epw,Antypas:2019qji} were able to place constraints on scalar DM models based on the variation of fundamental and physical constants.  One should keep in mind that, in general, dark stars can contribute to the overall DM abundance of the Universe. 
Gravitational lensing experiments such as MACHO~\cite{Alcock:2000ph} and EROS~\cite{Tisserand:2006zx} constrain the abundance of compact objects in
 the mass range $10^{-7}M_{\odot}\lesssim  M \lesssim 10 M_{\odot}$,  ($M_{\odot}$ being the solar mass)  to be less than $20\%$ of the total DM density of the Universe.

Star-like objects composed of ADM can be probed both by the aforementioned gravitational lensing studies but also by gravitational wave signals produced in the coalescence of such dark objects with black holes, other compact stars such as neutron stars, or among themselves~\cite{Giudice:2016zpa,Cardoso:2017cfl,Maselli:2017cmm,Maselli:2017vfi,Cardoso:2017cqb,Barack:2018yly,Cardoso:2019rvt,Pani:2019cyc}. Additional light signals can be also produced in particular scenarios where there is a portal that couples the dark sector to the SM one~\cite{Maselli:2019ubs}. However, if the dark star is sufficiently diffuse, as is the case for boson stars composed of ADM, gravitational waves produced in mergers of such objects are weak and alternative detection methods should be developed. One such method is via the use of high precision atomic clocks. The idea is simple.
Atomic clocks measure time by using specific atomic transitions. For microwave atomic clocks, the ticking of the clock is sensitively dependent on the fine structure constant as well as physical constants such as the masses of the electron and the quarks.  Whereas for optical atomic clocks, the ticking of the clock is only dependent on the fine structure constant \cite{Bize_2005,Derevianko:2013oaa}.  The passing of an atomic clock through a dilute ADM boson star could, under some conditions, change these parameters and cause the atomic clock to tick at a different rate than an atomic clock not covered by the boson star. Therefore, small de-synchronizations of atomic clocks located at different places on the Earth could indicate the passing of such a ADM boson star.  Clearly, once two  clocks located at different places are covered by the star, they again measure time with the same rate. The de-synchronization takes place only in the case where one clock is inside and another one outside the ADM boson star.

Previously, optical atomic clocks were able to reach a precision of $10^{-18}$ for the fractional frequency shift $\delta \omega / \omega$ ~\cite{PhysRevLett.104.070802,Bloom:2013uoa}, while more recently, a record precision of $10^{-19}$ \cite{PhysRevLett.120.103201} has been reached.  Microwave atomic clocks tend to be less sensitive, on the order of $10^{-16}$ \cite{Weyers_2018}, but it has been suggested that this precision can be improved to $10^{-17}\left(T \, (\text{K})/300\right)^2$ for certain alkali atoms \cite{Han_2019}.  These tools provide a new means to probe the existence of cosmological topological defects~\cite{Derevianko:2013oaa} or dilute ADM boson stars. In fact using data from the satellite born clocks of the Global Positioning System, the authors of \cite{Roberts:2017hla} managed to set new constraints on models where DM is in the form of topological defects.  Previous analyses \cite{Sikivie:2014lha,Krauss:2019lqo} have also shown how atomic clocks are affected if the DM is axionic in nature.

In this paper we present potential constraints that can be set on ADM boson stars (if these objects contribute to the DM relic abundance) using atomic clocks.  In particular, we investigate models where the bosons couple to the SM via a Higgs portal or through a photon portal providing a means to affect change in physical constants which determine the ticking of atomic clocks, such as the electron mass and the fine structure constant.  In this study, we focus on the effect of ADM boson stars on microwave atomic clocks when assuming the Higgs portal since, for this case optical atomic clocks are not sensitive probes as they are only affected by shifts in the fine structure constant.  Whereas, when we assume the photon portal, we focus on the effect of ADM boson stars on optical atomic clocks, since these tend to be more precise than microwave atomic clocks. Recently, the authors of \cite{Alonso-Alvarez:2019pfe} analyzed potential production mechanisms and evolution of light ADM, as well as the phenomenological consequences of the ADM coupling to the SM through a Higgs portal.  We point out that there have been a number of proposed scalar field DM models, relaxion \cite{Aharony:2019iad,Banerjee:2019epw} for example, and other phenomenological models \cite{Stadnik:2015kia,Hees:2018fpg} that couple directly to fermions or the electromagnetic field, and hence can cause variations in the masses of fermions and the fine structure constant.  These scalar field DM models can be constrained from various experiments and they can be probed by both optical and microwave atomic clocks.  The paper is organized as follows: In Sec. \ref{sec:boson_stars} we derive the density profile and mass-radius relation of ADM boson stars and we estimate the rate of events i.e., the frequency with which these objects pass through the Earth. In Sec. \ref{sec: ADSwAC}, we present the Higgs and photon portals that are responsible for shifting the timing of atomic clocks and we present updated constraints on the couplings involved in both portals. In Sec. \ref{sec: results} we identify the parameter space of ADM boson stars that can potentially be probed by future atomic clocks and finally we conclude in Sec. \ref{sec:conclusions}.

\section{Boson Stars} \label{sec:boson_stars}
As mentioned in the introduction star-like objects can be formed from bosonic DM which, at low temperatures, is in a BEC state. We analyze a $\phi^4$ theory for complex scalar fields, where the self-interaction potential is given by,
\begin{align} \label{eq:self_int_pot}
V(\phi \phi^*) = \pm\frac{\lambda}{4}(\phi \phi^*)^2,
\end{align}
where $\lambda$ is the self-coupling constant between bosons.  Here, the positive (negative) sign denotes repulsive (attractive) self-interactions.  One can find gravitationally bound systems composed of ADM subject to the self-interaction given by Eq. (\ref{eq:self_int_pot}) and analyze the collision rate of such systems with Earth based atomic clocks. 

\subsection{Density Profile}
In the case of repulsive self-interactions, one can solve the Einstein-Klein-Gordon (EKG) equation in order to derive the density and mass-radius profile of these objects (see e.g.~ \cite{Colpi:1986ye,Eby:2015hsq}).  In the case of attractive self-interactions, the relativistic effects are suppressed and it suffices to solve the Gross-Pitaevskii-Poisson (GPP) equations \cite{Boehmer:2007um,Chavanis:2011zi,Chavanis:2011zm}. In this paper, we focus on attractive self-interactions because they give objects that are more easily probed by atomic clocks. Namely, objects with smaller compactness (ratio of mass over radius), such as those composed of ADM with attractive self-interactions, have an increased probability of passing by the Earth, thus creating a de-synchronization in atomic clocks that are apart from each other.  On the contrary, repulsive self-interactions tend to create systems with higher compactness and, therefore, lower chances of passing by the Earth. Instead of exactly solving the GPP equations, an alternative variational method can be used~\cite{Chavanis:2011zi,Eby:2018dat}.  One can choose some variational ansatz for the wavefunction that characterizes matter in the boson star and minimizes the energy of the system. Taking attractive self-interactions corresponding to the negative sign in Eq. (\ref{eq:self_int_pot}) and assuming a non-relativistic expansion of the complex scalar field, 
\begin{align}
\phi = \frac{1}{\sqrt{2m}}e^{-i m t}\psi
\end{align}
where $m$ is the mass of the boson, the energy functional of the system is, 
\begin{equation}
E=\int d^3r \left (\frac{|\nabla \psi|^2}{2m} + \frac{1}{2} V_g |\psi|^2 - \frac{\lambda}{16 m^2}|\psi|^4\right),
\end{equation}
where $V_g$ is the self-gravitational potential which satisfies the Poisson equation,
\begin{align}
\nabla^2 V_g = 4\pi \frac{m^2}{M_P^2} |\psi|^2.
\end{align}
Here, $M_P$ is the Planck mass, and the wavefunction $\psi$ is normalized to the particle number $N$,
\begin{align}
\int d^3r |\psi|^2=N.
\end{align}

We choose an ansatz of the form \cite{Schiappacasse:2017ham},
\begin{equation} \label{eq: axionwave}
\psi_d\left(\frac{r}{\sigma_d}\right) = \sqrt{\frac{N}{7 \pi \, \sigma_d^3}} \left(1 + \frac{r}{\sigma_d}\right) \exp\left(-\frac{r}{\sigma_d}\right),
\end{equation}
where $\sigma_d$ is the dilute minimum energy solution to be found by minimizing the energy of the system.  In \cite{Eby:2018dat}, this ansatz was found to be an excellent approximation of numerical solutions for boson stars in the dilute region.  The minimization of the energy, with $N$ fixed, results in the dilute minimum energy solution, 
\begin{widetext}
\begin{equation} \label{eq: dilutesol}
\sigma_d = \frac{5376}{5373}\frac{M_P^2}{m^3 N} \left[1 + \sqrt{1 - \left(\frac{N}{N_\text{max}}\right)^2} \right] \simeq 10^{7}\, \text{km} \, \left(\frac{\mu\text{eV}}{m}\right)^3 \left(\frac{10^{57}}{N}\right)\left[ 1 + \sqrt{1 - \left(\frac{N}{N_\text{max}} \right)^2}\right],
\end{equation}
\end{widetext}
where $N_\text{max}$ is the maximum particle number beyond which no bound state solutions exist.  Therefore, for ADM boson stars subject to attractive self-interactions, the possible particle numbers are bounded from above by,
\begin{equation} \label{eq: Natt}
N \leq N_\text{max} = 10\frac{M_P}{m\sqrt{\lambda}} \simeq 10^{57} \,  \left(\frac{\mu\text{eV}}{m}\right)\sqrt{\frac{10^{-45}}{\lambda}}.
\end{equation}
The choice of ansatz given by Eq.~(\ref{eq: axionwave}) with the value of $\sigma_d$ that minimizes the energy, Eq.~(\ref{eq: dilutesol}), provides a good approximation for the wavefunction, $\psi_d$, of a  gravitationally bound dilute ADM boson star.  Note that in this regime, the self-gravitational energy plays an important role in the stability of the system \cite{Eby:2019ntd}.  The central density of the ADM boson star is given by,
\begin{widetext}
\begin{align} \label{eq:central_dens}
\rho(0) = m |\psi_d(0)|^2 = 2 m^2 |\phi(0)|^2 \simeq 10^{5} \, \text{GeV} \, \text{cm}^{-3}\left(\frac{m}{\mu\text{eV}}\right)\left(\frac{N}{10^{57}}\right)\left( \frac{10^{7} \, \text{km}}{\sigma_d} \right)^3.
\end{align}
\end{widetext}
We take the radius of the ADM boson star to be the radius inside which $99\%$ percent of the mass is contained, $R_{99}$, found via  
\begin{equation} \label{eq: R99int}
0.99 N = \int_0^{R_{99}} \left|\psi_d\left(\frac{r}{\sigma_d}\right)\right|^2 d^3 r,
\end{equation}
for a given particle number (or total mass).  For the ansatz chosen, $R_{99}$ is approximately equal to,
\begin{equation} \label{eq: R99}
R_{99}(N) \approx 5 \, \sigma_d.
\end{equation}

If all DM is in the form of such boson stars, it behaves as CCDM. If, however, only a fraction of DM bosons is in the form of boson stars, the self-interactions of the bosons have to obey well established limits from the bullet cluster and the ellipticity of galaxies (see  \cite{AmaroSeoane:2010qx,Eby:2015hsq} and the references therein).  For ADM, we assume $2 \rightarrow 2$ scattering between like charges subject to the interaction potential given by Eq. (\ref{eq:self_int_pot}).  In this case, the matrix element is $\mathcal{M} = i\lambda$ and the resulting cross-section is,
\begin{align}
\sigma(\phi\phi \rightarrow \phi\phi) = \frac{\lambda^2}{64 \pi m^2}.
\end{align}
Using the cross-section constraint obtained in \cite{Robertson:2016xjh}, we get an upper limit on the self-coupling $|\lambda|$,
\begin{equation} \label{eq: ALPcons}
\frac{\sigma}{m} \lesssim 2 \, \frac{\text{cm}^2}{\text{g}},
\qquad
|\lambda| \lesssim 10^{-21} \left( \frac{m}{\mu\text{eV}} \right)^{3/2}.
\end{equation}
In this study, we choose all of the local DM density to be composed of ADM boson stars.  In this case, the above constraint does not necessarily hold.  However, we find that for all possible parameter spaces obtained, the ADM self-coupling constants are well below the maximum value given by Eq. (\ref{eq: ALPcons}), and so we choose to keep the constraint when we search for the possible parameter space in Sec. \ref{sec: results}.

\subsection{Collision Rate}
We are interested in objects that can pass through the Earth at some minimum rate. As mentioned earlier, larger rates are achieved by objects that are relatively large and not massive. If these objects compose a component of DM, smaller masses correspond to larger number densities. Similarly, larger size increases the probability of passing through the Earth.
The scattering cross-section for collisions between either Earth or a detector on Earth and a boson star is (assuming non-relativistic speeds),
\begin{equation} \label{eq:xsec_BS_Earth}
\sigma \approx \pi (R_\text{target} + R_{99})^2,
\end{equation}
where $R_\text{target}$ is the radius of the target and $R_{99}$ is given by Eq. (\ref{eq: R99}).  
For all possible parameter spaces analyzed in Sec. \ref{sec: results}, the ADM boson stars have a size comparable to or much larger than the Earth.  Hence, the radius of the target in Eq. (\ref{eq:xsec_BS_Earth}) is taken to be the radius of the Earth $R_E$.  The mean free path for collisions is,
\begin{equation}
L = \frac{1}{n \sigma},
\end{equation}
where $n$ is the local number density of ADM boson stars which, assuming all DM is in the form of boson stars, is given by,
\begin{equation} \label{eq:numdens_BS}
n = \frac{\rho_\text{DM}}{m \, N} \simeq 10^{-17} \, R_E^{-3} \left(\frac{\mu\text{eV}}{m}\right) \left(\frac{10^{57}}{N}\right),
\end{equation}
where $\rho_\text{DM}\simeq 0.3 \, \text{GeV}/\text{cm}^3$ is the Earth's local DM density.  The frequency of collisions is then,
\begin{widetext}
\begin{equation} \label{eq: period_large}
f = \frac{v_E}{L} \simeq 10^{-3} \, \text{yr}^{-1} \left(\frac{\mu\text{eV}}{m}\right) \left(\frac{10^{57}}{N}\right) \left(\frac{R_{99}}{5\times 10^{7} \, \text{km}}\right)^2,
\end{equation}
\end{widetext}
where $v_E = 2.3\times10^2 \, \text{km} \, \text{s}^{-1}$ is the relative velocity between the  Earth and the ADM boson star.  Therefore, the collision frequency will not only depend on the number density of ADM boson stars, but also the boson mass $m$, and from Eqs. (\ref{eq: dilutesol}) and (\ref{eq: Natt}), the boson self-coupling constant $\lambda$.

\section{Probing Asymmetric Dark Stars with Atomic Clocks} \label{sec: ADSwAC}

\subsection{Higgs Portal and its Effect on Measured Parameters}
We are interested in ADM boson stars that can potentially be detected by atomic clocks. In this case, a portal that connects the dark sector and the SM one is needed.  In particular, we assume  that the DM sector communicates with the SM sector via a Higgs portal (see \cite{Piazza:2010ye,Stadnik:2016zkf,Flacke:2016szy}), i.e., there is a term in the Lagrangian of the form
\begin{equation} \label{eq: asymlagrangian}
\mathcal{L} = ... + \beta |\phi|^2 |H|^2,
\end{equation}
where $\beta$ is a positive constant.  We note that one can also include a term of the form $\phi |H|^2$ for real scalar fields, however in the case of a complex scalar field, this term is not invariant under the U(1) transformation $\phi \rightarrow e^{i\alpha} \phi$. Such a portal can open decay channels of the field $\phi$ to SM particles as long as $\phi$ is heavier than they are.  

The interaction between the ADM and the Higgs results in a shift of the Higgs vacuum expectation value (VEV) given by,
\begin{align}\label{eq:HiggsVEV}
v = v_\text{ew}\sqrt{1 - \frac{2 \beta v_\phi^2}{m_h^2}} \approx v_\text{ew}\left(1 - \frac{\beta v_\phi^2}{m_h^2}\right)
\end{align}
where $v_\text{ew}$ is the VEV of the Higgs for $\beta = 0$, $v_\phi$ is the nonzero expectation value of $\phi$, $m_h$ is the Higgs mass, and the last equality holds for $\beta v_\phi^2 \ll m_h^2$.  In many  cases throughout this paper, this assumption will hold, and from this point we will take $v \approx v_\text{ew}$ unless explicitly stated otherwise.  Notice that in order that the Higgs obtains a nonzero VEV, it must be true that $\beta v_\phi^2 < m_h^2$.  Given that the ADM density in the early universe was large, for a given value of $\beta$ it could be possible that the Higgs VEV vanishes.  Because of this, we assume that the ADM forms at a time such that the Higgs VEV always exists once it forms sometime before BBN.

Notice that the interaction between the ADM and the Higgs adds to the mass term of the $\phi$ field, so that after the Higgs acquires a VEV, the effective mass of $\phi$ is given by,
\begin{align}
m^2 \approx m_{\phi,\text{bare}}^2 + \beta v_\text{ew}^2.
\end{align}
It should be noted that for any parameter space corresponding to an observable frequency shift of an atomic clock, the bare mass of $\phi$ squared must be fine tuned such that the appropriate mass $m$ is obtained.  The self-coupling constant $\lambda$ must also be fine-tuned due to loop corrections of the $|\phi|^2 |H|^2$ coupling.  

If it is assumed that the complex scalar field $\phi$ is bound in a dilute boson star, it is related to the dilute wavefunction and central density of the boson star as given by Eq. (\ref{eq:central_dens}).  Therefore, the presence of the star induces an effective change in the mass of the electron through the Higgs portal of the following form
\begin{equation} \label{eq: mass_shift}
m_e= \frac{y_e v}{2} \approx m_e^{\text{bare}} \left(1 - \frac{\beta v_{\phi}^2}{m_h^2}\right),
\end{equation}
where $m_e^\text{bare}$ is by definition the electron mass in the absence of any medium and $y_e$ is the Yukawa coefficient for the electron.  Note that, depending on the sign of $\beta$, the effective mass can be larger or smaller than the bare mass of the electron $m_e^{\text{bare}}$.  In this study, we take $v_\phi$ to be the central value of $\phi$ inside the boson star given by,
\begin{widetext}
\begin{align} \label{eq:vphi}
v_\phi \equiv |\phi(0)| = \sqrt{\frac{\rho(0)}{2m^2}}
\simeq 1 \, \text{MeV} \left\{\left(\frac{N}{10^{57}}\right)\left(\frac{\mu\text{eV}}{m}\right)\left(\frac{10^7\, \text{km}}{ \sigma_d}\right)^3\right\}^{1/2}.
\end{align}
\end{widetext}

\subsection{Photon Portal and its Effect on Measured Parameters}
One can also couple the SM to the dark sector through a photon portal \cite{Stadnik:2015kia}.  In this case, the Lagrangian has a term of the form,
\begin{align}
\mathcal{L} = ... + \frac{g}{4}|\phi|^2 F^2
\end{align}
where the coupling constant $g$ can be positive or negative.  This interaction causes a shift in the fine structure constant given by,
\begin{align}
\alpha = \alpha_0 \left(\frac{1}{1 - g \, v_\phi^2}\right)\approx \alpha_0 \left(1 + g \, v_\phi^2 \right)
\end{align}
where $v_\phi$ is given by Eq. (\ref{eq:vphi}) and the last equality holds when $g v_\phi^2 \ll 1$.  Again, we take this assumption to hold throughout this paper, unless explicitly stated otherwise.

\subsection{Frequency Shift of Atomic clocks}
For atomic clocks in general, the change in the counting of the clock follows~\cite{Derevianko:2013oaa}\begin{equation} \label{eq: freq_shift}
\frac{\delta \omega}{\omega}=\frac{\delta V}{V},
\end{equation}
where
\begin{equation} \label{eq: freq}
V=\alpha^{K_{\alpha}}\left (\frac{m_q}{\Lambda_\text{QCD}}\right )^{K_q} \left (\frac{m_e}{m_p} \right )^{K_{e/p}},
\end{equation}
and $\alpha$, $m_q$, $m_e$ and $m_p$ are the the fine structure constant, the quark mass, the electron mass, and the proton mass, respectively. $\Lambda_\text{QCD}$
is the scale of QCD and the $K_i$ are appropriate constants for the corresponding quantities $i$ depending on the particular atom used in the atomic clock.

For a typical microwave atomic clock~\cite{Derevianko:2013oaa}, $K_{\alpha} \simeq 2$, $K_q \simeq - 0.09$ and $K_{e/p} = 1$. Given the portal of Eq.~(\ref{eq: asymlagrangian}), $\alpha$ remains unchanged. The change in the mass of the quarks makes very little contribution to $\delta\omega$ due to the small factor 0.09 and since most of the mass of the proton does not come from the mass of the quarks, this is also a tiny contribution. Therefore, in our setup the overwhelming  contribution to $\delta\omega$ comes from the change of electron mass.  Assuming the Higgs portal, as an ADM boson star passes through the Earth, the mass of the electron is shifted due to the nonzero value of $v_{\phi}$ via Eq.~(\ref{eq: mass_shift}). In this case, the shift in frequency is given by,
\begin{align} \label{eq: freq_micro}
\left(\frac{\delta \omega}{\omega}\right)_\text{Higgs}  \simeq \frac{\delta m_e}{m_e}\simeq \frac{\beta v_\phi^2}{m_h^2} \simeq 10^{-20} \left(\frac{\beta}{10^{-10}}\right) \left(\frac{v_\phi}{\text{MeV}}\right)^2
\end{align}
Notice that for the benchmark parameters, this frequency shift is several orders of magnitude smaller than the detectable frequency shift of the most precise microwave atomic clocks currently.  

Assuming, instead, the photon portal the effect of a passing ADM boson star is a shift in frequency given by,
\begin{align}\label{eq:freq_phot}
\left(\frac{\delta \omega}{\omega}\right)_\text{Photon}  \simeq \frac{\delta \alpha}{\alpha}\simeq 10^{-19} \left(\frac{g}{10^{-13} \, \text{GeV}^{-2}}\right) \left(\frac{v_\phi}{\text{MeV}}\right)^2
\end{align}
Notice that both microwave and optical atomic clocks will approximately exhibit this frequency shift.  In this case, and given the much better precision of optical atomic clocks, we focus on the shift of optical atomic clocks for this portal.

It is apparent that larger boson star densities correspond to larger values of $v_\phi$, which consequently create larger shifts in the mass of the electron and therefore larger $\delta\omega$ shifts in the clock timing. The reader should recall however that usually larger densities are achieved in heavier stars which have smaller collision rates with the Earth. Therefore, the class of boson stars that can be probed are those that have large enough $\delta\omega$ so that the change in timing is detectable while at the same time the collision rate remains relatively high. We explore different values for the DM self-coupling constant as well as for the coefficients $\beta$ and $g$. As discussed earlier, we assume that all DM is composed of ADM boson stars, and hence the DM self-coupling constant is not necessarily constrained by Eq. (\ref{eq: ALPcons}).  However, in scanning the parameter space, we find that the possible self-coupling constants do, in fact, satisfy this constraint, and so we choose to search the parameter space with this constraint satisfied.  The Higgs coupling constant $\beta$ and the photon coupling constant $g$ are also subjected to different types of constraints as demonstrated in the next subsection.

\subsection{Bounds on Higgs and Photon Couplings} \label{sec:  Higgsbounds}
An upper bound on the Higgs coupling constant $\beta$ in Eq. (\ref{eq: asymlagrangian}) can be found from the observed constraint on the branching fraction of invisible Higgs decays.  From \cite{Kouvaris:2014uoa}, the rate for the invisible Higgs decay is given by,
\begin{align}
\Gamma(h \rightarrow \phi \phi) \approx \frac{\beta^2 v_\text{ew}^2}{8 \pi m_h} \left( 1 - \frac{4 m^2}{m_h^2} \right)^{1/2}.
\end{align}
Recent measurements from the CMS collaboration \cite{Sirunyan:2018owy} give an upper constraint on the branching fraction of invisible Higgs decays of $19\%$ at $95\%$ CL.  For $\Gamma(h \rightarrow \text{SM}) = 4.1 \, \text{MeV}$ and taking $m \ll m_h$, an upper constraint on $\beta$ is found to be, 
\begin{equation} \label{eq: invhconst}
\beta \lesssim 10^{-2}.
\end{equation}

It has been shown that constraints can be placed on the change in the Fermi constant throughout the evolution of the universe \cite{Scherrer:1992na}.  Assuming both the Fermi constant and fermion masses change as a result of the ADM density, the ratio of the Fermi constant at the start of BBN, $G_F^\text{BBN}$ to the Fermi constant today, $G_F^0$, has an upper limit given by,
\begin{align}
\frac{G_F^\text{BBN}}{G_F^0} = \frac{1 - \beta \rho_{\text{DM},0}^\text{avg}/(2 m^2 m_h^2)}{1 - \beta \rho_{\text{DM,BBN}}^\text{avg}/(2m^2 m_h^2)} < 1.01
\end{align}
where $\rho_{\text{DM},0}^\text{avg} = 1.3\, \text{keV}\,\text{cm}^{-3}$ is the average DM density of the universe today and $\rho_{\text{DM,BBN}}^\text{avg}$ is the average DM density of the universe at a temperature of $1 \, \text{MeV}$.  Taking the redshift corresponding to this temperature to be $z_\text{BBN} = 4 * 10^{9}$, the average DM density of the universe was $\sim 10^{29}$ times greater than the value today, and hence, the shift in the Fermi constant could have been significant.  The constraint on the change in the Fermi constant constrains the Higgs coupling constant to be \cite{Alonso-Alvarez:2019pfe},
\begin{align}\label{eq:beta_BBN}
\beta \lesssim 2 \times 10^{-10}\left(\frac{m}{\mu\text{eV}}\right)^2\left(\frac{1.3 \, \text{keV}{\text{cm}}^{-3}}{\rho_\text{DM,0}^\text{avg}}\right).
\end{align}

Notice that for this benchmark mass $m$, the constraint from invisible Higgs decays (Eq. (\ref{eq: invhconst})) is more stringent.  However for masses $m\lesssim 10^{-2} \, \text{eV}$, the constraint from BBN (Eq. (\ref{eq:beta_BBN})) starts to become comparable.  In the next section, we will show that because of these constraints, the possible frequency shifts of microwave atomic clocks from boson stars subject to the Higgs portal are several orders of magnitude smaller than the currently detectable frequency shift.  However, the BBN bound will change if we assume that the ADM forms after BBN \cite{Blum:2014vsa}.  It has been theorized that certain classes of CDM can form between BBN and the time of matter radiation equality \cite{Sarkar:2014bca}.  Alternatively, one can assume that the ADM formed after BBN as a decay product of another particle that has no direct coupling to the Higgs.  In these cases, the possible parameter space may open up, and we leave such analyses for future work.

The most strenuous constraints for the photon coupling constant in the possible parameter spaces of the next section are from BBN and energy loss of supernovae.  As with the Fermi constant, the change in the fine structure constant from the time of BBN until now is also constrained \cite{Stadnik:2015kia},
\begin{align}\label{eq:g_BBN}
g \lesssim 8 \times 10^{-14}\, \text{GeV}^{-2} \left(\frac{m}{\mu\text{eV}} \right)^2,
\end{align}
while energy loss constraints of supernovae give
\begin{align}\label{eq:g_SN}
g \lesssim 10^{-7} \, \text{GeV}^{-2}.
\end{align}
We show, in the next section, that in taking account both constraints on the photon coupling constant, we obtain an available parameter space that gives a detectable frequency shift assuming the collision frequencies with Earth are small ($f \sim 10^{-2} \, \text{yr} $).

\section{Results} \label{sec: results}
In this section we would like to identify the ADM boson star parameter space that can be probed by Earth based atomic clocks. The constraint on $\beta$ from inverse Higgs decays (Eq. (\ref{eq: invhconst})) and from BBN (Eq. (\ref{eq:beta_BBN})), the constraint on the condensate particle number for boson stars with attractive self-interactions (Eq. (\ref{eq: Natt})), and the constraint on the boson self-coupling constant (Eq. (\ref{eq: ALPcons})), provide boundaries for an available parameter space to scan when assuming the Higgs portal.  Again, we stress that this last constraint is not necessary since we assume that all DM is in the form of ADM boson stars.  However, in searching the available parameter space without satisfying this constraint, we find that all possible solutions do, in fact, keep this constraint satisfied.  When we assume the photon portal, we scan the available parameter space assuming the constraints on the photon coupling (Eqs. (\ref{eq:g_BBN}) and (\ref{eq:g_SN})), as well as the last two constraints mentioned above (Eqs. (\ref{eq: Natt}) and (\ref{eq: ALPcons})).

We scan the available parameter space to find solutions for which the rate of collisions between  boson stars and the Earth given by Eq. (\ref{eq: period_large}) is $f \geq f_\text{min}$, and the size of the boson star is $R_{99} < 10^{10} \, \text{km}$.  This last constraint arises due to the fact that solutions with $R_{99} \geq 10^{10} \, \text{km}$ will take a year or more to completely pass through a detector on Earth.  We also take the constraint that the frequency shift from Eqs. (\ref{eq: freq_micro}) and (\ref{eq:freq_phot}) is $\delta \omega / \omega \geq  (\delta \omega / \omega)_\text{min}$.  Finally, the solutions found satisfy the condition that the ADM boson stars found locally do not significantly overlap i.e.,
\begin{align}
\rho(0) \geq \rho_\text{DM},
\end{align}
where $\rho(0)$ is the central density of the ADM boson star given by Eq. (\ref{eq:central_dens}).

The rate of collisions depends on $\lambda$, $m$, and $N$ (the number of particles composing the star), while the induced fractional frequency shift $\delta \omega / \omega$ depends on $\lambda$, $m$, $N$, and $\beta$ or $g$. We take the constraints on $\lambda$, $N$, $\beta$, and $g$ as described previously. We scan the parameter space by varying the relevant parameters within the ranges $10^{-22} \, \text{eV} \leq m \leq 10^6 \, \text{eV}$, $10^{-100}\lambda_\text{max}<\lambda<\lambda_\text{max}$, $0.01N_{\text{max}}<N<N_\text{max}$, and $10^{-50}\beta_\text{max}<\beta<\beta_{\max}$ (assuming the Higgs portal) or $10^{-50}g_\text{max}<g<g_{\max}$ (assuming the photon portal). From this scan, we identify the parameter space that can provide a frequency of collision $f \geq f_\text{min}$ with an induced $\delta\omega/\omega\geq (\delta\omega/\omega)_\text{min}$ provided that all DM is in the form of these ADM boson stars, and that the radius and central density of the boson stars satisfy the constraints as described above.

\subsection{Higgs portal}
\begin{figure}[t] 
\centering
\includegraphics[width=\linewidth,height=225pt]{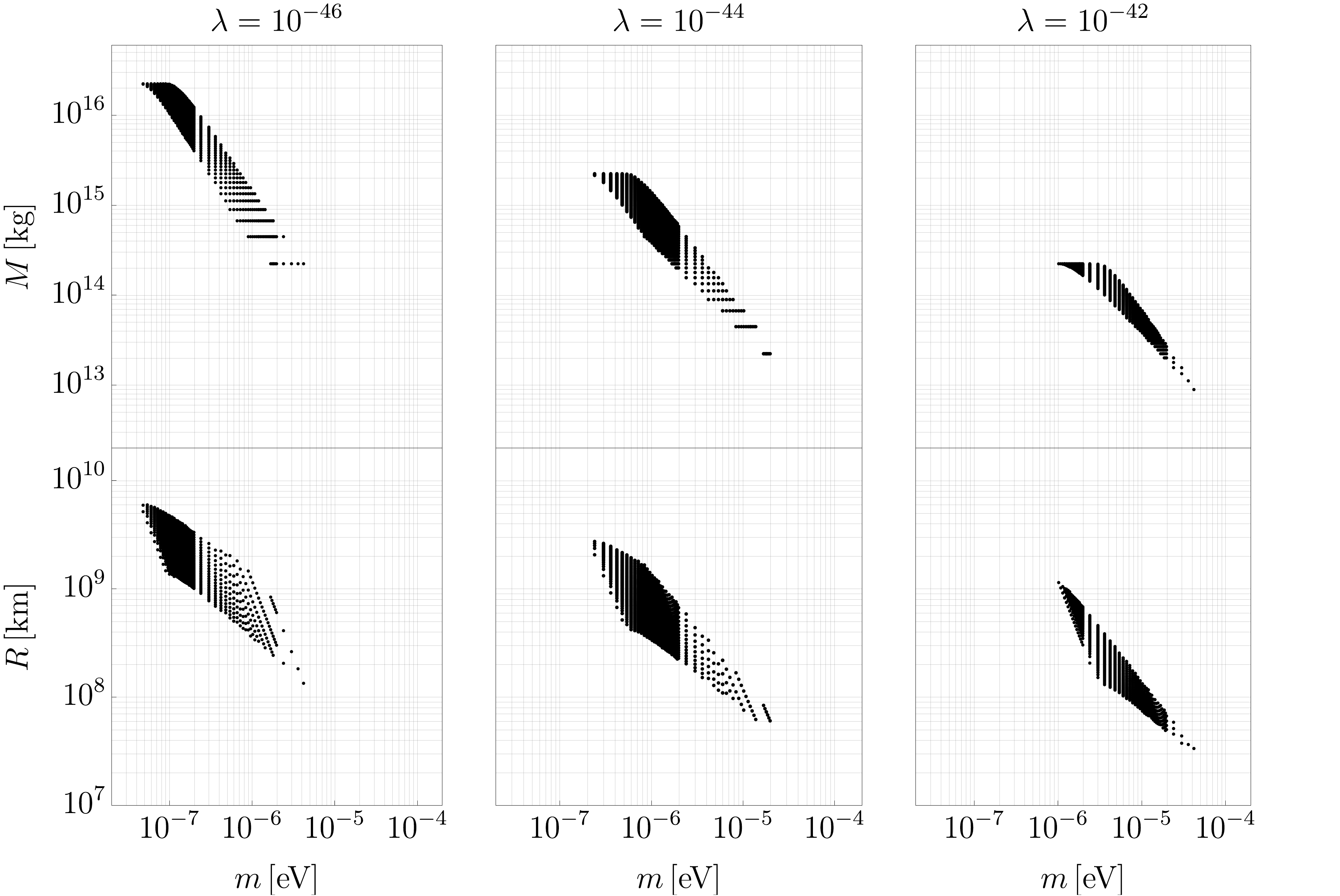} \\[\abovecaptionskip]
\caption{Total mass $M$ and radius $R$ of the boson star subject to the Higgs portal vs. particle mass $m$ for which $\beta = 10^{-2}$, $\delta \omega/\omega \geq 10^{-18}$, the frequency of collisions between the ADM boson star and the detector is $f \geq 1 \, \text{yr}^{-1}$, and the self-coupling of the ADM is $\lambda = 10^{-46}$ (left panel), $\lambda = 10^{-44}$ (middle panel), and $\lambda = 10^{-42}$ (right panel).  Notice that this value of $\beta$ is excluded from BBN constraints if the ADM is assumed to have formed before BBN.} \label{fig:paramspace_Higgs}
\end{figure}

\begin{figure}[t]
\centering
\includegraphics[width=\linewidth,height=225pt]{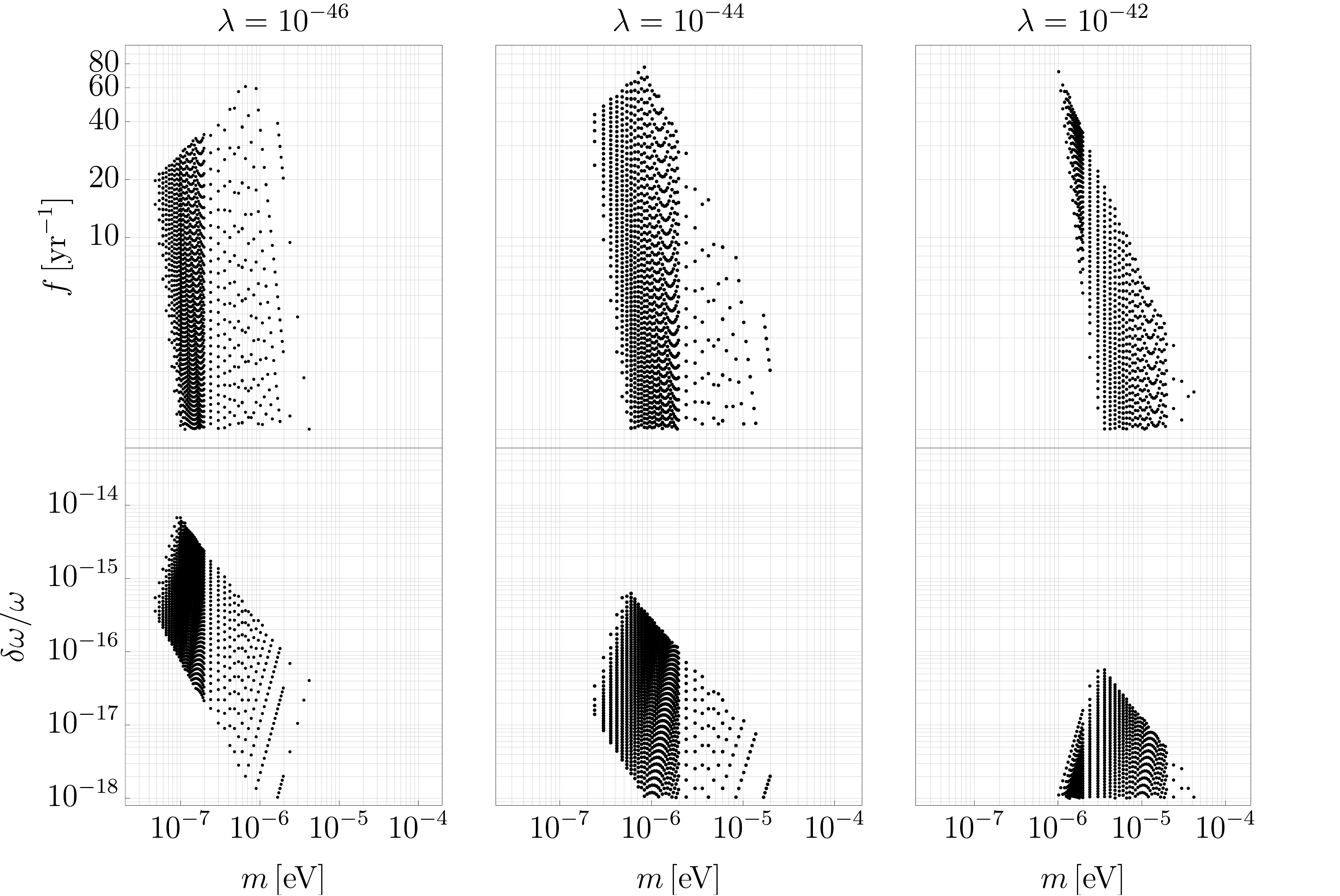} \\[\abovecaptionskip]
\caption{Collision frequency $f$ and frequency shift $\delta \omega / \omega$ vs. particle mass $m$ for the parameter space in Fig. \ref{fig:paramspace_Higgs}.} \label{fig:fw_Higgs}
\end{figure}

\begin{figure}[t]
\centering
\includegraphics[width=\linewidth,height=250pt]{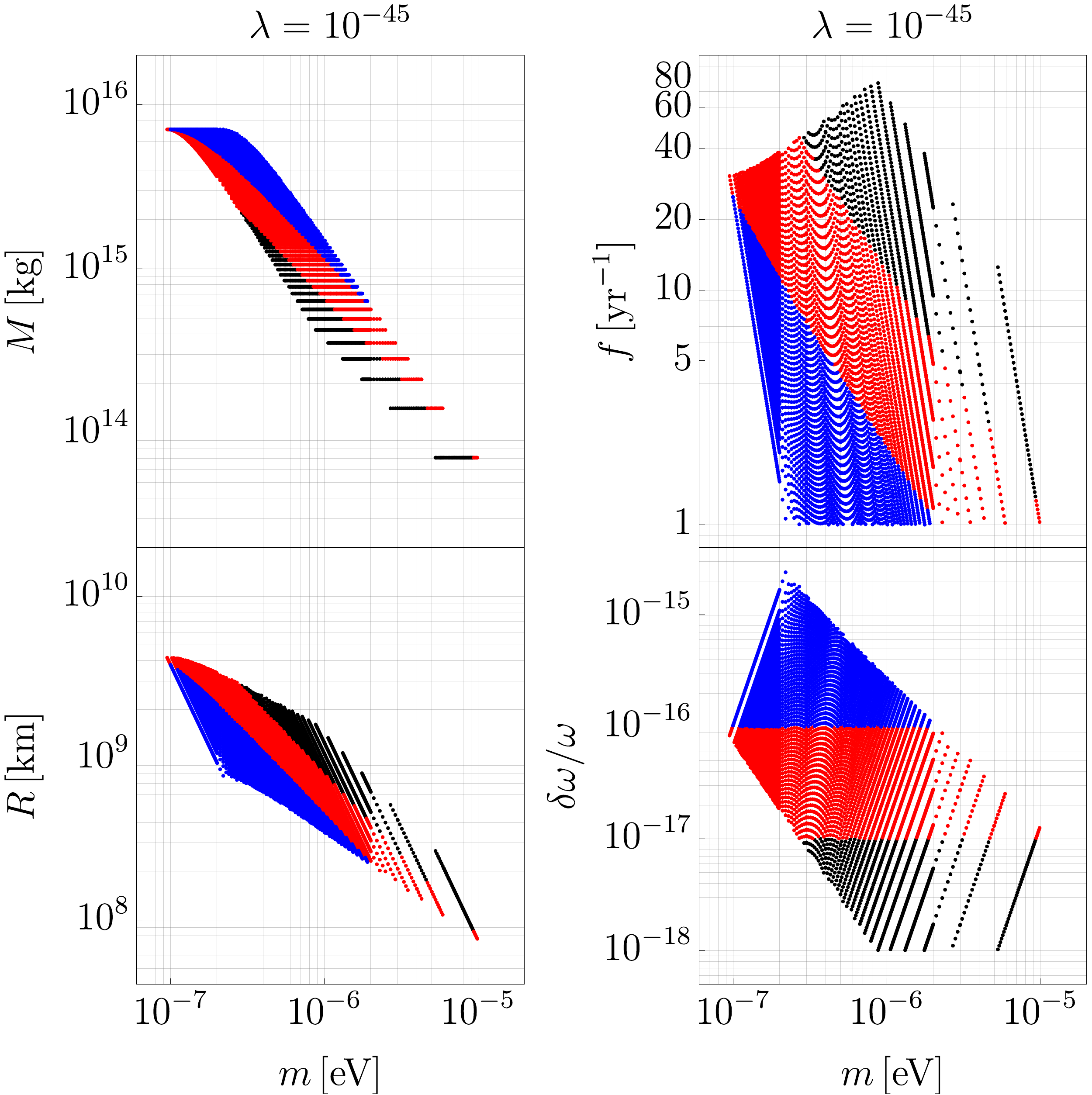} \\[\abovecaptionskip]
\caption{Total mass $M$ (top left), radius $R$ (bottom left), collision frequency $f$ (top right), and frequency shift $\delta \omega / \omega$ (bottom right) of the ADM boson star vs. particle mass $m$ for which $\beta = 10^{-2}$, $\lambda = 10^{-45}$, and the frequency of collision between the boson star and detector is $f \geq 1 \, \text{yr}^{-1}$.  Blue dots corresponds to $\delta \omega / \omega \geq 10^{-16}$, red dots to $10^{-16} > \delta \omega / \omega \geq 10^{-17}$, and black dots to $10^{-17}> \delta \omega / \omega \geq 10^{-18}$.  Notice that this value of $\beta$ is excluded from BBN constraints if the ADM is assumed to have formed before BBN.} \label{fig:paramspace_45_Higgs}
\end{figure}

\begin{figure}[t]
\centering
\includegraphics[width=\linewidth,height=250pt]{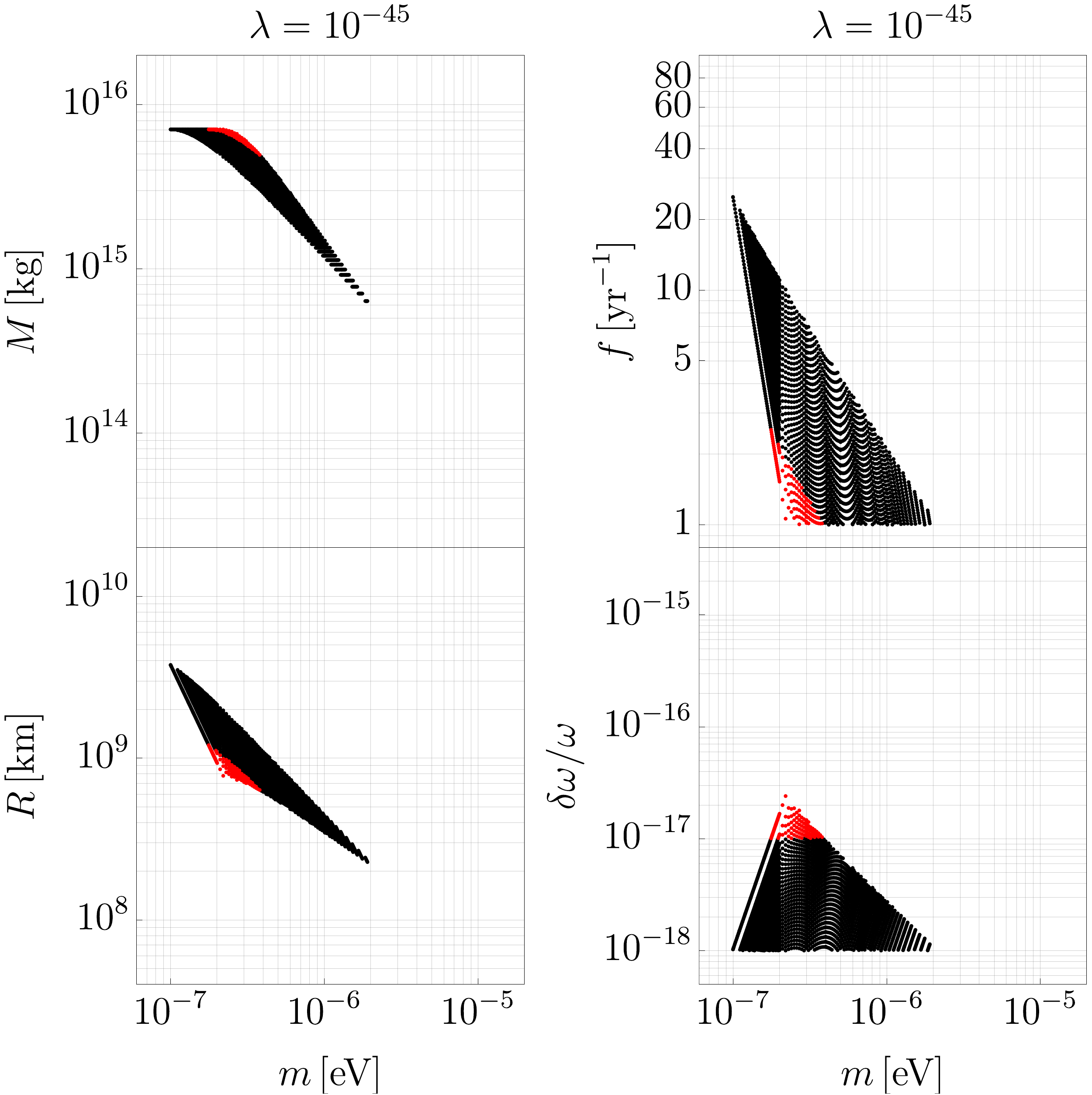} \\[\abovecaptionskip]
\caption{Total mass $M$ (top left), radius $R$ (bottom left), collision frequency $f$ (top right), and frequency shift $\delta \omega / \omega$ (bottom right) of the ADM boson star vs. particle mass $m$ for which $\beta = 10^{-4}$, $\lambda = 10^{-45}$, and the frequency of collision between the boson star and detector is $f \geq 1 \, \text{yr}^{-1}$.  Red dots corresponds to $10^{-16} > \delta \omega / \omega \geq 10^{-17}$ and black dots to $10^{-17}> \delta \omega / \omega \geq 10^{-18}$.  Notice that this value of $\beta$ is excluded from BBN constraints if the ADM is assumed to have formed before BBN.} \label{fig:paramspace_45_Higgs_comp}
\end{figure}

We show the following parameter space for a boson star subject to the Higgs portal obtained when the BBN constraint on the Higgs coupling constant (Eq. (\ref{eq:beta_BBN})) is negligible.  We emphasize that this constraint will change if the ADM forms long after BBN and it may be possible that the invisible Higgs decay is the most strenuous constraint.  We leave such analyses for future work, and show the results that one could obtain when the BBN constraints  are negligible.  In Fig. \ref{fig:paramspace_Higgs} we show the mass and radius of the boson star as a function of the DM mass where $\delta\omega/\omega\geq10^{-18}$ with a rate of events  larger than one per year, after having chosen three different values of $\lambda$ and having fixed $\beta=0.01$. In Fig. \ref{fig:fw_Higgs} we show how the aforementioned parameter space is distributed in terms of $\delta\omega/\omega$ and rate of events, while in Fig. \ref{fig:paramspace_45_Higgs} we show with different colors which part of the parameter space requires atomic clock sensitivity $10^{-16}$, $10^{-17}$ or $10^{-18}$ in order to detect the passing of such a dilute boson star from the Earth.  In Fig. \ref{fig:paramspace_45_Higgs_comp}, we show how the available parameter space changes when the Higgs coupling constant $\beta$ is decreased by two orders of magnitude.  One can see that the available parameter space decreases and that no events give a fractional frequency shift greater than $10^{-16}$.

Notice that for the parameter space shown in Figs. \ref{fig:paramspace_Higgs}, \ref{fig:fw_Higgs}, \ref{fig:paramspace_45_Higgs}, and \ref{fig:paramspace_45_Higgs_comp}, the values of $\beta = 10^{-2}$ and $\beta = 10^{-4}$ are constrained from BBN.  We see then, from Eq. (\ref{eq: freq_micro}) that if we take a value of $\beta$ as constrained from BBN, the frequency shift induced is several orders of magnitude less than the most precise microwave atomic clocks even for rare events.  Conversely, we can attempt to increase the frequency shift by increasing $v_\phi$.  From Eq. (\ref{eq:vphi}), one can see that this can be done by either increasing the number of particles, or by decreasing the particle mass or size of the boson star.  Notice from Eqs. (\ref{eq: dilutesol}), (\ref{eq: Natt}), and (\ref{eq: period_large}), there is a delicate balance that one must achieve between the free parameters of the boson mass $m$ and the self-coupling constant $\lambda$ in order to satisfy all constraints and obtain a detectable frequency shift for microwave atomic clocks.  Because of all the necessary constraints, we find that for boson stars subject to the Higgs portal, the frequency shift induced is several orders of magnitude smaller than the currently detectable frequency shift for microwave atomic clocks.  However, as noted above, if the BBN constraint can be shifted due the formation of the ADM after BBN, the parameter space could open considerably.  Also, one can open the parameter space by assuming a smaller minimum collision frequency with the Earth.  For example, one can satisfy all constraints, assuming the ADM forms before BBN, if the minimum collision frequency is lowered to $f \sim 10^{-5} \, \text{yr}^{-1}$.

Several comments are in order here. First, it is obvious that further improvements to microwave atomic clock sensitivities will lead to an extension of the parameter space probed within this class of models as is apparent from Fig. \ref{fig:paramspace_45_Higgs}. For all of the parameter space shown, we have assumed that these dilute boson stars make up $100\%$ of the DM relic abundance in our galaxy. More parameter space can be probed by atomic clocks if one relaxes this condition. If boson stars compose a smaller fraction of DM, part of the parameter space can still be probed as long as the rate of events remains sufficient. This can happen for example in cases where the dilute boson star is large yet it makes up a small fraction of DM because the probability of passing through the Earth can still remain high. In addition, the clocks of the GPS network have been collecting data for more than 10 years and therefore, the same technique used in \cite{Roberts:2017hla} can be used to probe boson stars that could have a rate of events of $\sim 0.1/\text{year}$ if the accuracy of the GPS clocks improve in the near future.

We can also define the difference in the induced fractional frequency shift between two atomic clocks as,
\begin{align}
\frac{\delta \omega}{\omega}\Big|_\text{rel} = \frac{\delta \omega}{\omega}\Big|_1 - \frac{\delta \omega}{\omega}\Big|_2
\end{align}
where the induced fractional frequency shift at clocks $1$ and $2$ can be determined as a function of time,
\begin{align}
\frac{\delta \omega}{\omega}(t) = \frac{\beta}{m_h^2} \frac{\left|\psi_d\left(\frac{v_E|t|}{\sigma_d}\right)\right|^2}{2m},
\end{align}
where $\psi_d$ is given by Eq. (\ref{eq: axionwave}) and $\sigma_d$ is given by Eq. (\ref{eq: dilutesol}).  

\begin{figure}[t]
\centering
\includegraphics[width=\linewidth,height=250pt]{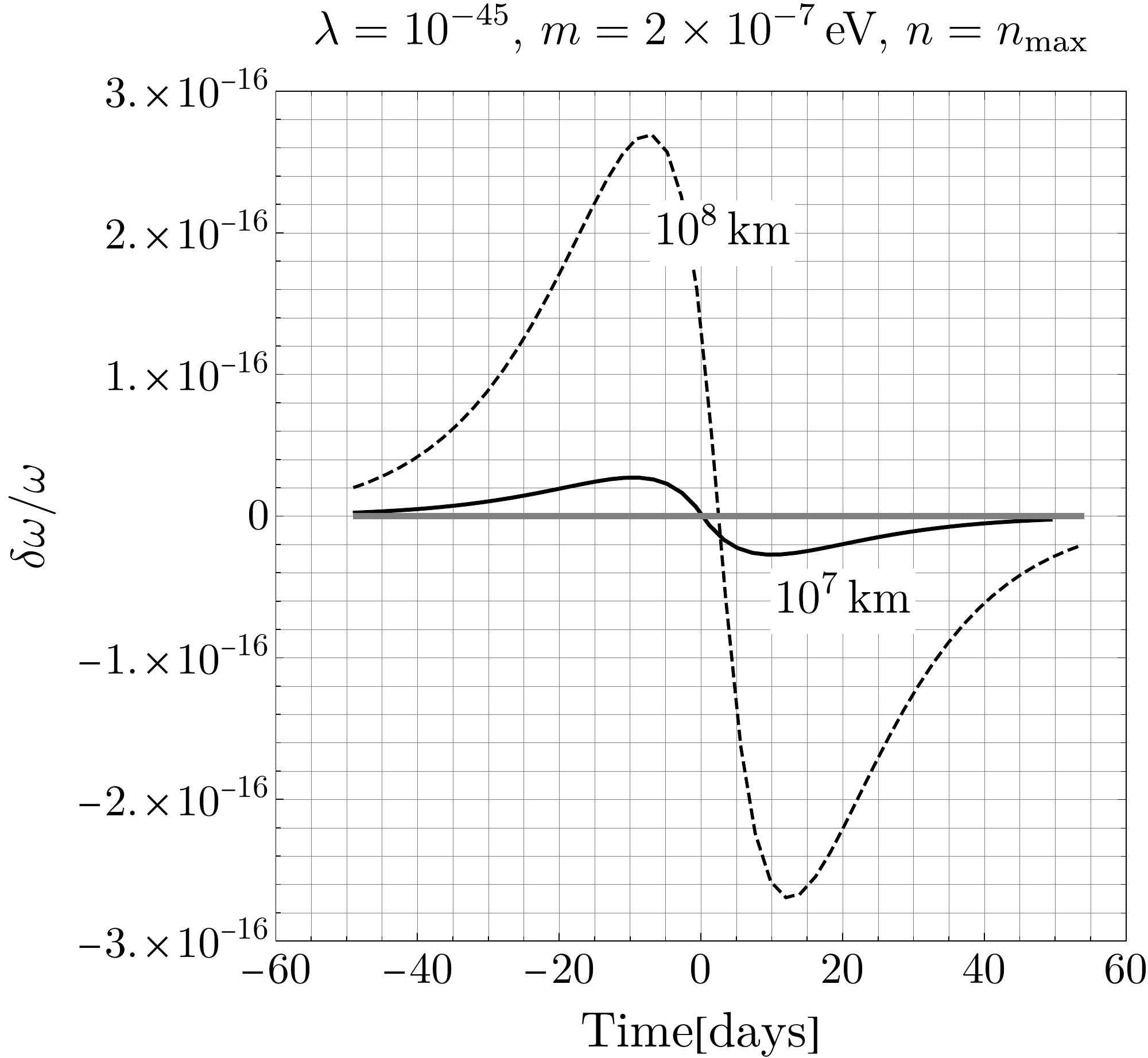} \\[\abovecaptionskip]
\caption{The relative induced fractional frequency shift $\delta \omega/\omega|_{\text{rel}}$ vs. time as two synchronized microwave atomic clocks pass through a boson star.  The boson star is taken to have parameters $\lambda = 10^{-45}$, $m = 2\times 10^{-7} \, \text{eV}$, $n = n_\text{max}$, and $\beta = 10^{-2}$ which results in $M \sim 10^{16} \, \text{kg}$, $R_{99} \sim 10^{9} \, \text{km}$, and $f \sim 2 \, \text{yr}^{-1}$.  The minimum time here is taken to be when the edge of the boson star starts to pass through the first clock, while the maximum time is taken to be the time at which the boson star fully passes through the second clock.  The dashed line corresponds to a distance of $10^8 \, \text{km}$ between the detectors while the thick line corresponds to a distance of $10^7 \, \text{km}$.  The filled in region corresponds to $\left|\delta \omega/\omega|_{\text{rel}} \right| \leq 10^{-18}$.  Notice that this value of $\beta$ is excluded from BBN constraints if the ADM is assumed to have formed before BBN.}\label{fig:signal_shift_rel_Higgs}
\end{figure}

\begin{figure}[t] 
\centering
\includegraphics[width=\linewidth,height=250pt]{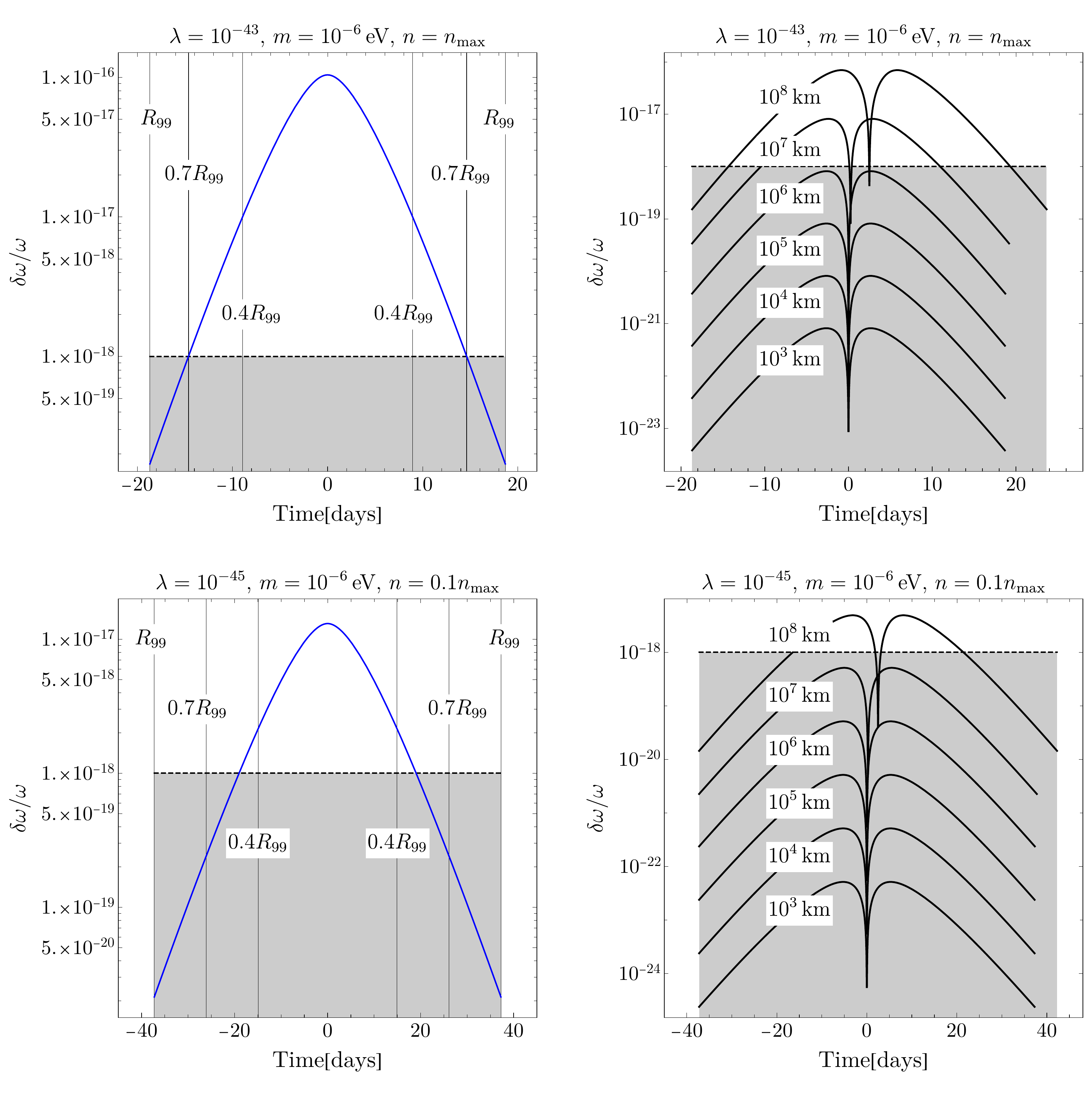} \\[\abovecaptionskip]
\caption{The absolute magnitude of the induced fractional frequency shift $\delta \omega/\omega$ vs. time as two synchronized atomic clocks pass through a boson star.  The left panel shows the signal for two synchronized microwave and optical atomic clocks while the right panel shows the signal for two synchronized microwave atomic clocks.  The top panel shows the signal for $\lambda = 10^{-43}$, $m = 10^{-6} \, \text{eV}$, and $n = n_\text{max}$ which results in a boson star with $M \sim 10^{15} \, \text{kg}$, $R_{99} \sim 10^{8} \, \text{km}$, and $f \sim 2 \, \text{yr}^{-1}$, while the bottom panel shows the signal for $\lambda = 10^{-45}$, $m=10^{-6}\, \text{eV}$, and $n = 0.1 n_\text{max}$ which results in a boson star with $M \sim 10^{15} \, \text{kg}$, $R_{99} \sim 10^{9} \, \text{km}$, and $f \sim 10 \, \text{yr}^{-1}$.  For all plots, it is assumed that $\beta = 10^{-2}$.  The gridlines in the left panels show the position of the microwave atomic clock inside the boson star, while the labels in the right panels show the distance between the two microwave atomic clocks.  Notice that this value of $\beta$ is excluded from BBN constraints if the ADM is assumed to have formed before BBN.} \label{fig:signal_shift_Higgs}
\end{figure}

Fig. \ref{fig:signal_shift_rel_Higgs} shows $(\delta \omega/\omega|_\text{rel})$ between two microwave atomic clocks for different distances between the clocks.  Fig. \ref{fig:signal_shift_Higgs} shows $(\delta \omega/\omega|_\text{rel})$ between two synchronized optical and microwave \cite{JaredEvans,Kim:2008,Bergeron:2016} (left panel) and the absolute value of $(\delta \omega/\omega|_\text{rel})$ between two synchronized microwave (right panel) atomic clocks given two sets of parameter space.  Since optical atomic clocks are not sensitive to the passing of these particular boson stars, the difference in the induced fractional frequency shift is either always positive or always negative depending on how one defines the difference between the fractional frequency shifts.  Notice that for both of these plots, the Higgs coupling constant is $\beta = 10^{-2}$, which is excluded from BBN constraints for this mass range.  If it is assumed that the ADM forms before BBN, then the frequency shifts would decrease by the appropriate orders of magnitude.  

One can see from the left panel of Fig. \ref{fig:signal_shift_Higgs} that, if it is assumed that one can neglect the BBN constraint for the Higgs coupling constant, for a given set of ADM parameters, a fractional frequency shift that is greater than $\delta \omega / \omega \sim 10^{-18}$ is induced between a given pair of synchronized microwave and optical atomic clocks (independently of the distance between the clocks).  In this case, there is some chance of observing passing ADM boson stars with future microwave atomic clock sensitivities, assuming one can neglect the BBN bounds on the Higgs coupling constant.  However, one can see from the right panel of Fig. \ref{fig:signal_shift_Higgs} that the prospect of observing passing ADM boson stars with two synchronized microwave atomic clocks is more impractical.  Because the size of the systems that correspond to the available parameter space of Figs. \ref{fig:paramspace_Higgs}, \ref{fig:fw_Higgs}, and \ref{fig:paramspace_45_Higgs} tend to be $\mathcal{O}(10^8 - 10^{10}) \, \text{km}$, and the induced fractional frequency shift $\mathcal{O}(10^{-18} - 10^{-14})$, the distance between the two synchronized microwave atomic clocks must be large in order to obtain a difference in the signal that is greater than $10^{-18}$.  Of course, smaller systems will result in synchronized microwave atomic clocks that can be put closer together while still getting an observable signal, however these systems will collide with the Earth less often. 

Note that measurements made with the atomic clocks are subject to the uncertainty principle (i.e. $\delta \omega \delta t \geq 1$ where $\delta \omega$ is the frequency shift and $\delta t$ is the transient time of the boson star).  For a typical microwave frequency of $10^{10} \, \text{Hz}$, the transient time of the boson star must satisfy $\delta t \geq 10^{-10} \left(\delta \omega/\omega\right)^{-1} \, \text{s}$.  Notice from Fig. \ref{fig:signal_shift_rel_Higgs}, that the fractional frequency shift is $\delta \omega / \omega \simeq 10^{-16}$ for a transient time of the boson star $\delta t \simeq 3 \times 10^{6} \, \text{s}$.  In this case, the uncertainty principle is just satisfied.

\subsection{Photon portal}
\begin{figure}[t] 
\centering
\includegraphics[width=\linewidth,height=250pt]{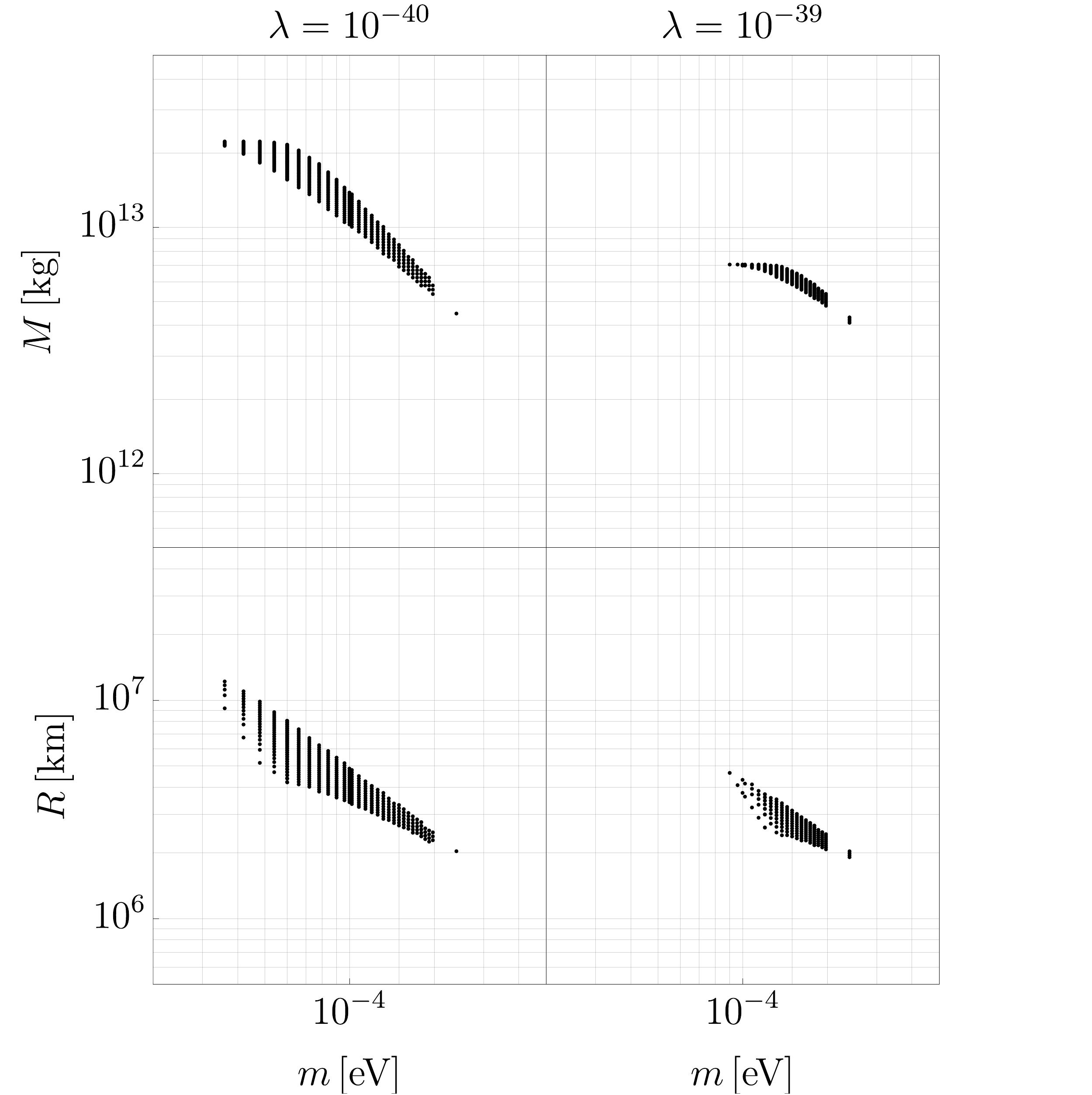} \\[\abovecaptionskip]
\caption{Total mass $M$ and radius $R$ of the boson star subject to the photon portal vs. particle mass $m$ for which $g = 10^{-10} \, \text{GeV}^{-2}$, $\delta \omega/\omega \geq 10^{-20}$, the frequency of collisions between the ADM boson star and the detector is $f \geq 10^{-2} \, \text{yr}^{-1}$, and the self-coupling of the ADM is $\lambda = 10^{-40}$ (left panel) and $\lambda = 10^{-39}$ (right panel).} \label{fig:paramspace_20}
\end{figure}

\begin{figure}[t]
\centering
\includegraphics[width=\linewidth,height=250pt]{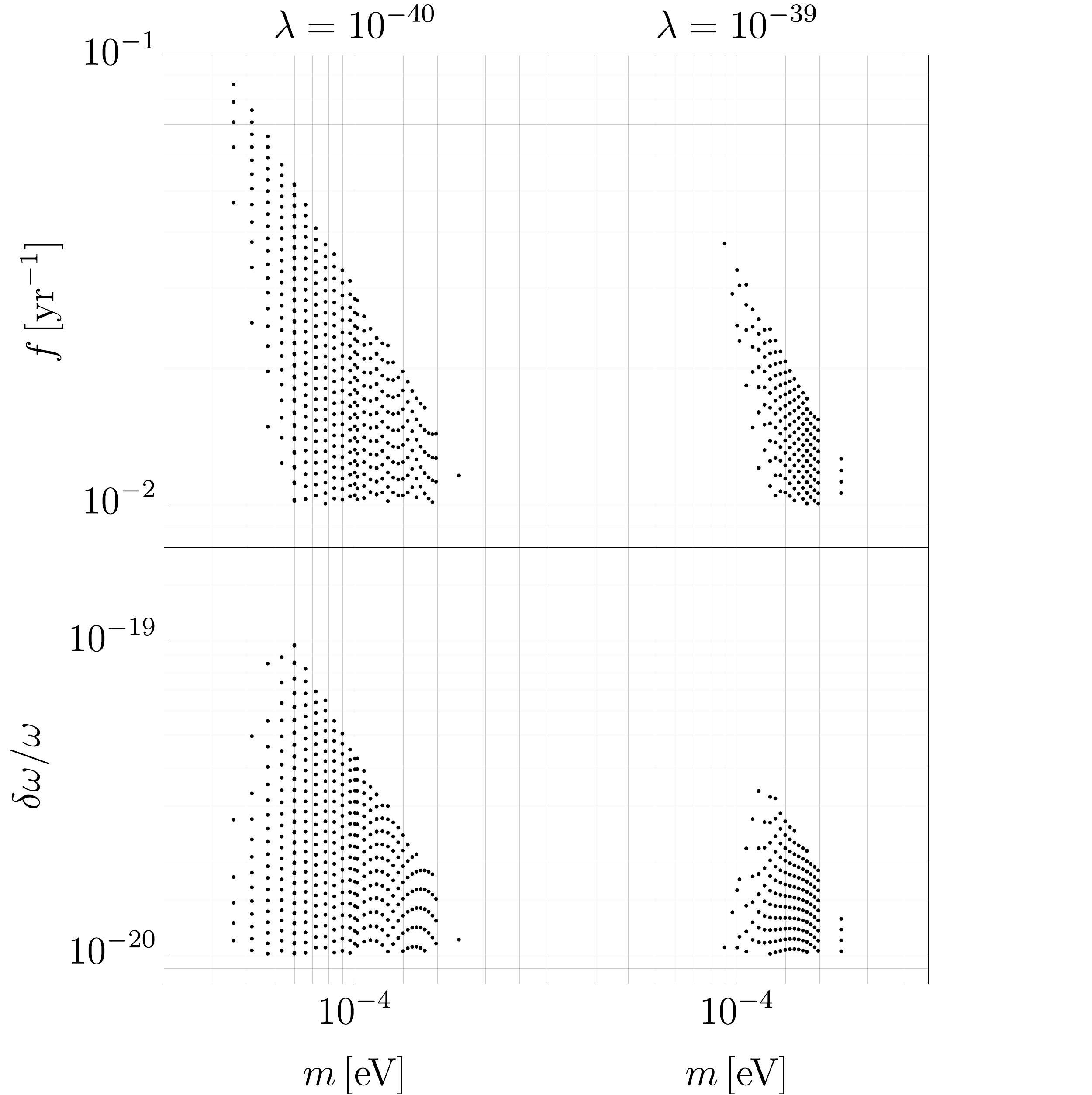} \\[\abovecaptionskip]
\caption{Collision frequency $f$ and frequency shift $\delta \omega / \omega$ vs. particle mass $m$ for the parameter space in Fig. \ref{fig:paramspace_20}.} \label{fig:fw_20}
\end{figure}

\begin{figure}[t] 
\centering
\includegraphics[width=\linewidth,height=250pt]{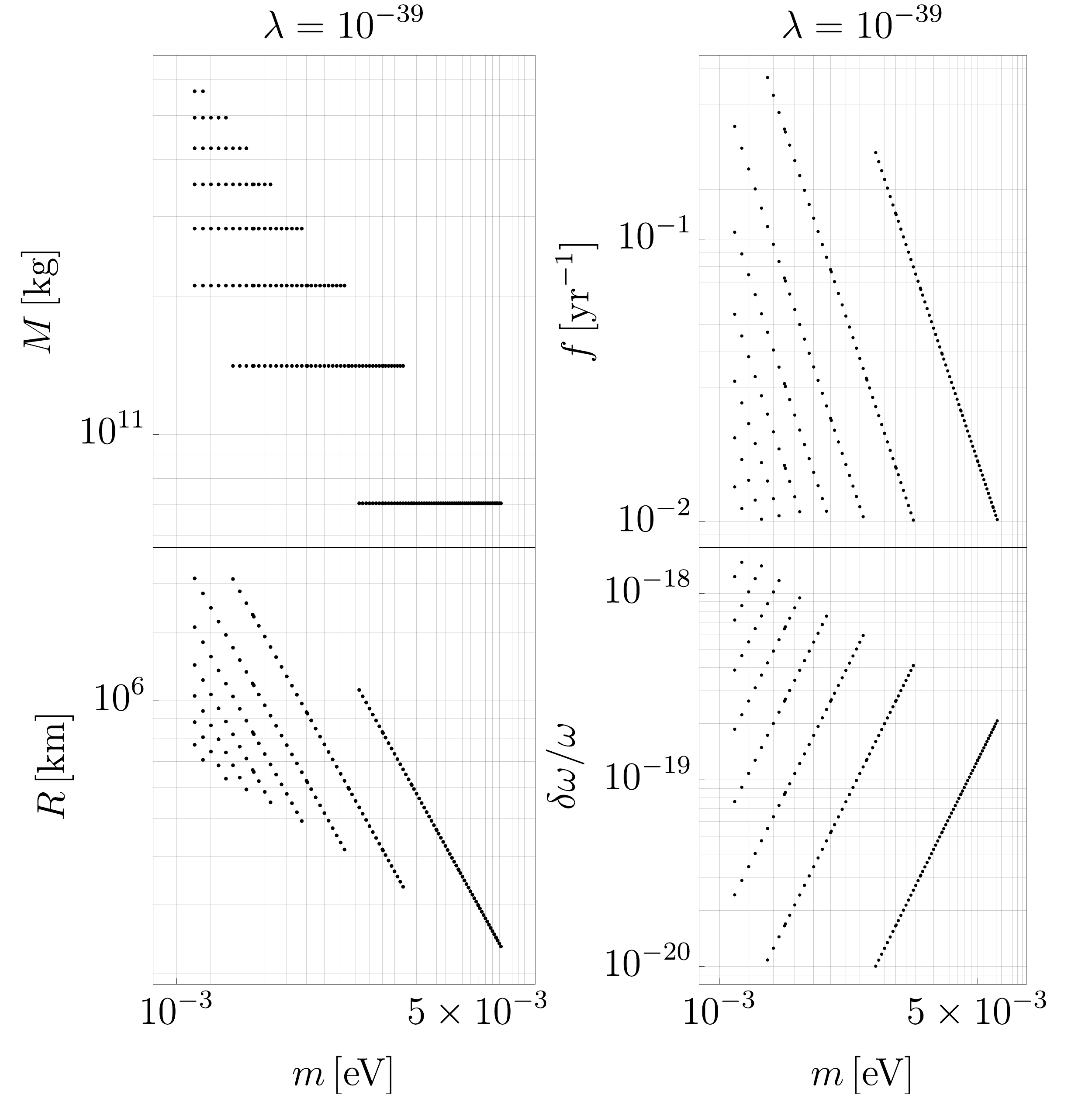} \\[\abovecaptionskip]
\caption{otal mass $M$, radius $R$, collision frequency $f$, and frequency shift $\delta \omega / \omega$ for a boson star subject to the photon portal vs. particle mass $m$ for which $g = 10^{-7} \, \text{GeV}^{-2}$, $\delta \omega/\omega \geq 10^{-20}$, the frequency of collisions between the ADM boson star and the detector is $f \geq 10^{-2} \, \text{yr}^{-1}$, and the self-coupling of the ADM is $\lambda = 10^{-39}$.} \label{fig:paramspace_20_new}
\end{figure}

\begin{figure}[t]
\centering
\includegraphics[width=\linewidth,height=250pt]{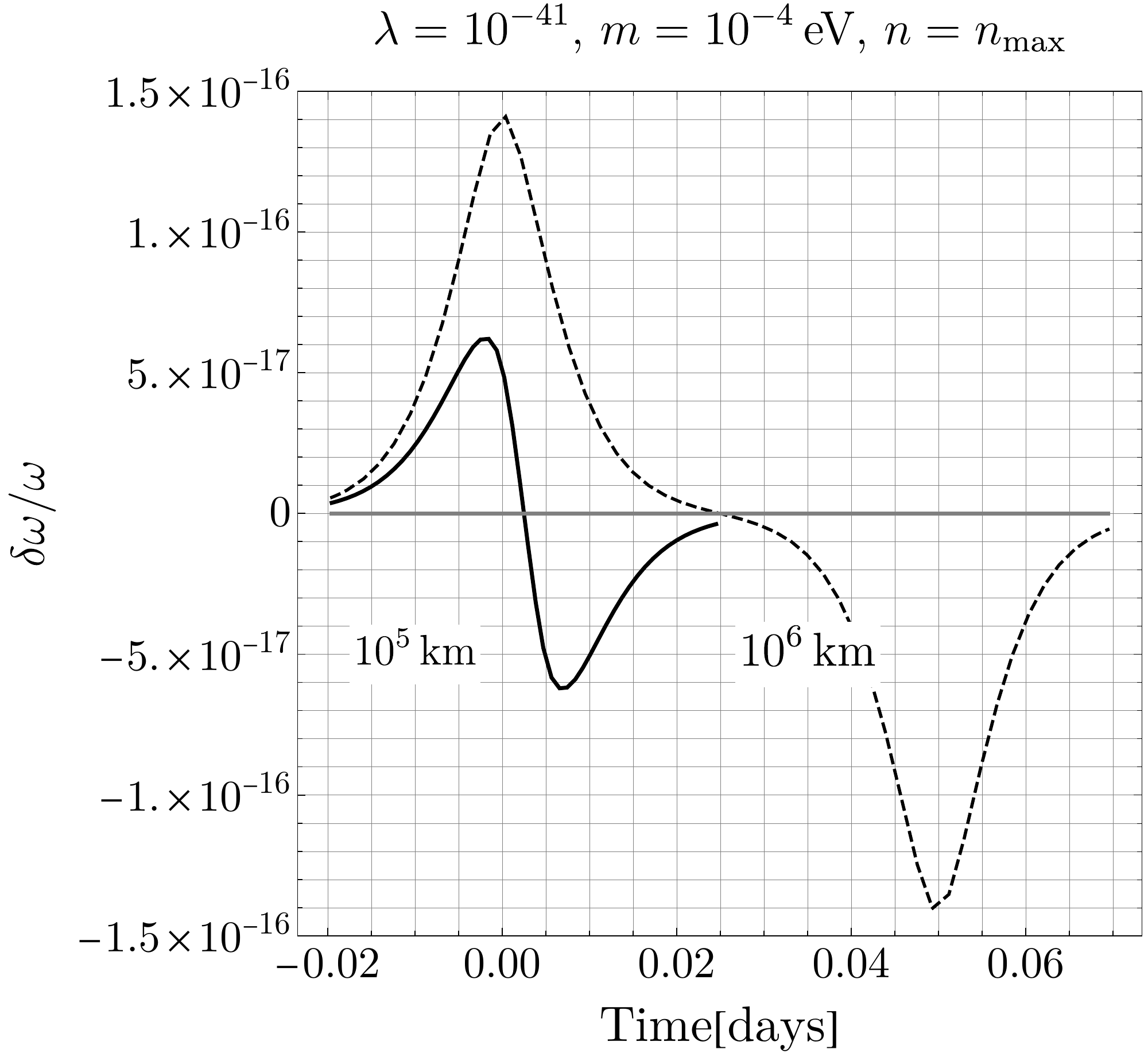} \\[\abovecaptionskip]
\caption{The relative induced fractional frequency shift $\delta \omega/\omega|_{\text{rel}}$ vs. time as two synchronized optical atomic clocks pass through a boson star subject to the photon portal.  The boson star is taken to have parameters $\lambda = 10^{-41}$, $m = 10^{-4} \, \text{eV}$, $n = n_\text{max}$, and $g = 10^{-10}\,\text{GeV}^{-2}$ which results in $M \sim \times 10^{14} \, \text{kg}$, $R_{99} \sim 10^{5} \, \text{km}$, and $f \sim 10^{-5} \, \text{yr}^{-1}$.  The minimum time here is taken to be when the edge of the boson star starts to pass through the first clock, while the maximum time is taken to be the time at which the boson star fully passes through the second clock.  The dashed line corresponds to a distance of $10^6 \, \text{km}$ between the detectors while the thick line corresponds to a distance of $10^5 \, \text{km}$.  The filled in region corresponds to $\left|\delta \omega/\omega|_{\text{rel}} \right| \leq 10^{-20}$}\label{fig:signal_shift_rel_new}
\end{figure}

In Fig. \ref{fig:paramspace_20} we show the mass and radius of the boson star subject to the photon portal as a function of the DM mass where $\delta\omega/\omega\geq10^{-20}$ with a frequency of collision events $f\geq 10^{-2}\,\text{yr}$, after having chosen two different values of $\lambda$ and having fixed $g=10^{-10}\,\text{GeV}^{-2}$ in order to satisfy both the BBN and supernova constraints.  In Fig. \ref{fig:fw_20} we show how the aforementioned parameter space is distributed in terms of $\delta\omega/\omega$ and rate of events.  We also show how the available parameter space changes when the photon coupling constant $g$ is increased by three orders of magnitude in Fig. \ref{fig:paramspace_20_new}.  Fig. \ref{fig:signal_shift_rel_new} shows $(\delta \omega/\omega|_\text{rel})$ between two optical atomic clocks for different distances between the clocks.  Notice that the parameter space for which all constraints are satisfied correspond to rare events.

Finally, as described above, the uncertainty principle should hold throughout the transient time of the boson star.  For a typical optical frequency of $10^{14} \, \text{Hz}$, the transient time of the boson star must satisfy $\delta t \geq 10^{-14} \left(\delta\omega/\omega\right)^{-1}\,\text{s}$.  From Fig. \ref{fig:signal_shift_rel_new}, one can see that the uncertainty relation holds for a fractional frequency shift $\delta \omega/\omega \gtrsim 5 \times 10^{-17}$.  In this case, though, the collision frequency between the boson star and the Earth is very rare, $f \simeq 10^{-5} \, \text{yr}^{-1}$.

\section{Conclusions} \label{sec:conclusions}
In this paper, we entertain the possibility that dark matter is entirely composed of light asymmetric dark matter with attractive self-interactions that has collapsed in dilute formations. If the dark sector communicates with the visible sector via a Higgs or photon portal, the passing of such a dilute object through the Earth can induce a small change in physical constants like the mass of the electron or the fine structure constant.  Due to the fact that dark matter in boson stars is in a BEC state, the nonzero expectation value of the boson field creates an extra contribution to the mass of the electron or the fine structure constant. Since the timing frequency of atomic clocks depends on these parameters, a clock that finds itself embedded in the boson star as the latter crosses the Earth, measures time at a different rate compared to a clock that remains, at that time, outside the star. We search the parameter space for a class of dilute boson stars subject to a photon or Higgs portal where conventional techniques such as gravitational waves from mergers, gravitational lensing and direct dark matter detection fail.  We demonstrate that, taking into account all constraints on the Higgs and photon coupling constants, the induced frequency shift of both microwave and optical atomic clocks is several orders of magnitude smaller than the currently detectable frequency shift for these instruments or the events are rare.

In particular, we assume that the complex scalar field composing the asymmetric dark matter boson stars has a quadratic coupling to the Higgs or to the photon.  We discuss the constraints that the dark matter self-coupling, Higgs coupling constant, and photon coupling constant must satisfy.  We then scan the available parameter space subject to these constraints.  Additionally, we set the constraints that the frequency of collisions between a boson star and the Earth and the induced fractional frequency shift due to the shift in the electron mass or fine structure constant are greater than some minimum values, that the boson stars do not overlap, and that the radius of the boson stars are small enough to pass the Earth within one year. For both the Higgs and photon portals, we find that the induced frequency shifts are several orders of magnitude smaller than the currently detectable frequency shifts for microwave and optical atomic clocks.  However, it may be possible that the ADM forms after BBN, in which case, the constraint on the Higgs coupling constant will change and may open up some available parameter space.  We also see more available parameter space by taking a smaller minimum frequency of collisions between the ADM boson stars and the Earth.  For the photon portal, we begin to obtain some available parameter space satisfying all constraints when assuming one collision every one hundred years.  We stress that even if the accuracies of atomic clocks improve considerably in the near future, such probes of astrophysical objects are still subject to the uncertainty principle which can diminish the available parameter space.

\section*{Acknowledgments}

We thank Prof. Kelin Gao for the numerous discussions we had on atomic clocks and the vital information he provided to us on the setup and precision of the atomic clocks of his lab.  We are grateful to Joshua Eby, Jared Evans, and Peter Suranyi for many valuable discussions.  L.S. and L.C.R.W. thank the University of Cincinnati Office of Research Faculty Bridge Program for funding through the Faculty Bridge Grant.  L.C.R.W. thanks the Aspen Center for Physics, which is supported by the National Science Foundation grant PHY-1607611, where some of the research was conducted. C.K. is partially funded by 
the Danish National Research Foundation, grant number DNRF90, and by the Danish Council 
for Independent Research, grant number DFF 4181-00055.

\appendix
\section{Possible Additional Constraints}
Constraints can be placed on the Higgs coupling constant, $\beta$, in Eq. (\ref{eq: asymlagrangian}) from fifth-force experiments if a nonzero expectation value of $\phi$ exists at the location of the experiment.  One way $\phi$ can obtain an expectation value is if $\phi$ gets its mass from the Higgs and the Higgs coupling constant $\beta$ is negative \cite{Cosme:2018nly}.  In this case, the expectation value is different from that defined in Eq. (\ref{eq:vphi}), and we leave the search for the possible parameter space corresponding to these systems for future studies.  Another way fifth-force experiments can constrain $\beta$, is if the field $\phi$ is assumed to form a condensate around or inside the Earth, Sun, etc. \cite{Banerjee:2019epw,Nelson:2018xtr,Horowitz:2019pru}.  In this case, fifth-force experiments on the Earth will always be affected by the $\phi$ expectation value given by Eq. (\ref{eq:vphi}).  In this study, we show the constraints on $\beta$ that arise given that the $\phi$ field has a nonzero expectation value that effects the fifth-force experiments.  However, we do not take these constraints into account for our calculations as we assume that the ADM boson stars are not bound to the Earth as a halo and refer to \cite{Banerjee:2019epw} for such discussions.

From \cite{Piazza:2010ye,Finkbeiner:2008qu,Grinstein:2018ptl}, the presence of $\phi$, with a mass $m$, induces an interaction between two massive bodies with a potential,
\begin{align}
V(r) = - \frac{m_1 m_2}{r} \left( \frac{\alpha}{M_P} \right)^2 e^{-m r},
\end{align}
where $m_i$ is the mass of the $i$-th body and $\alpha$ is a coupling constant given by,
\begin{align} \label{eq:fifthforce1}
\alpha = g_{hNN} \frac{\sqrt{2} M_P}{m_N} 
	\epsilon.
\end{align}
Here, $m_N$ is the nucleon mass, 
$\epsilon$ is the mixing parameter which is proportional to the Higgs coupling constant $\beta$ in Eq. (\ref{eq: asymlagrangian}),
\begin{align} \label{eq:higgsmixing}
\epsilon \approx \frac{\beta \, v_\phi \, v_\text{ew}}{m_h^2},
\end{align}
where we again neglected terms of $\mathcal{O}(\beta^2)$ and $g_{hNN}$ is the coupling of the Higgs to nucleons given by
\begin{align}
g_{hNN} = \sum_{q=u,d,s,c,b,t} \langle N| \bar{q} q | N \rangle g_{hqq} = \sum_{q=u,d,s,c,b,t} \frac{f^N_{Tq} m_N}{v_\text{ew}}.
\end{align}

\begin{figure}[t]
  \centering
  \includegraphics[width=\linewidth,height=200pt]{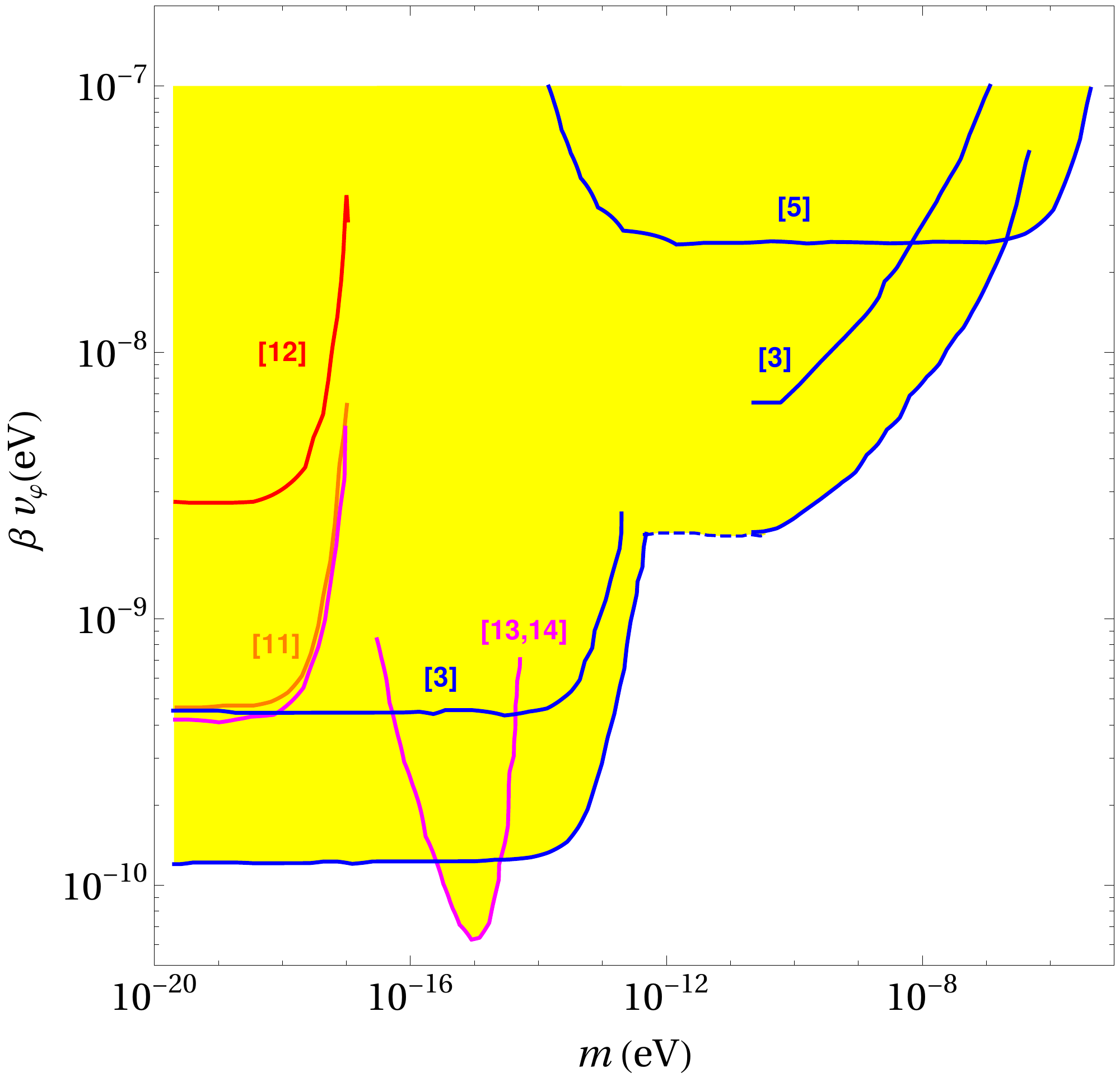}
  \includegraphics[width=\linewidth,height=200pt]{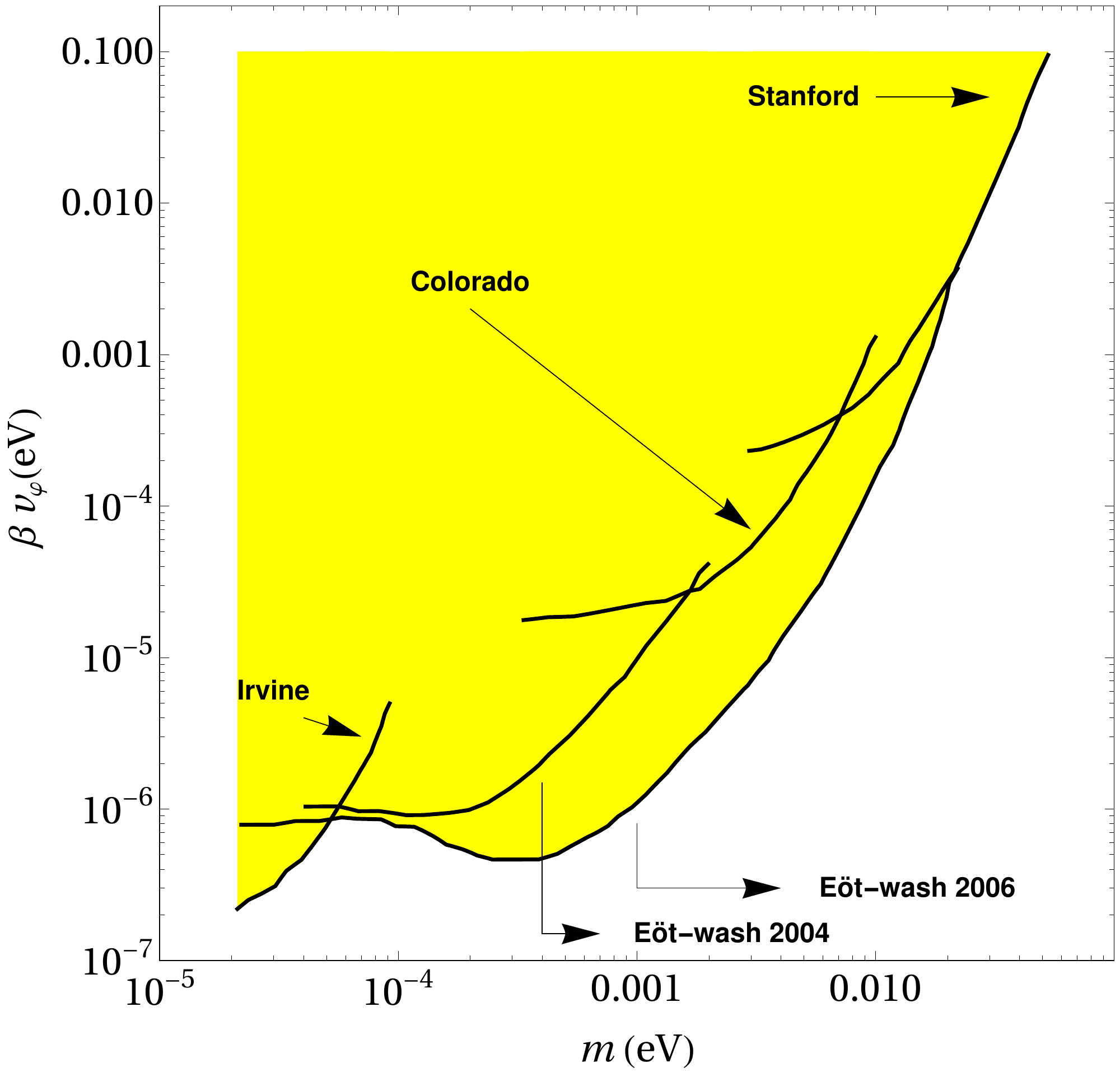}
  \caption{Constraints on the product of the Higgs coupling $\beta$ with $v_\phi$ given by Eq. (\ref{eq:vphi}) from gravitational inverse-square law tests \cite{Piazza:2010ye,PhysRevLett.100.041101,
Bertotti:2003rm,PhysRevLett.98.021101}.  The reference labels in the left panel correspond to those described in \cite{PhysRevLett.100.041101}, while the reference labels in the right panel correspond to those described in \cite{PhysRevLett.98.021101}.}  \label{fig:fifthforcelin}
\end{figure}

Here, $f_{Tq}^N$ are the nucleon parameters \cite{Jungman:1995df,Cheng:1988im,Gasser:1990ce,Finkbeiner:2008qu} and it has been used that,
\begin{align}
\langle N | \bar{q} q | N \rangle = f^N_{Tq} m_N / m_q 
\qquad
g_{hqq} = m_q / v_\text{ew}.
\end{align} 
For protons and neutrons, the couplings are $g_{hpp} \approx 0.3776 \, m_p / v_\text{ew}$ and $g_{hnn} \approx 0.3755 \, m_n/v_\text{ew}$, respectively.  Taking an average of these two couplings, Eq. (\ref{eq:fifthforce1}) becomes,
\begin{align} \label{eq:fifthforce2}
\alpha \sim 10^{5} \left( \frac{\beta \, v_\phi}{\text{eV}} \right).
\end{align}
The value of $\alpha^2$ is constrained by the aforementioned fifth-force experiments~\cite{Piazza:2010ye,PhysRevLett.100.041101,Bertotti:2003rm,PhysRevLett.98.021101} and from Eq. (\ref{eq:fifthforce2}) one can draw constraints on $\beta \, v_\phi$ as depicted in Fig.~\ref{fig:fifthforcelin}.

If the U(1) symmetry of the Lagrangian is unbroken, then $\phi$ is protected from decays into two photons.  However, if the U(1) charge of $\phi$ is not conserved, one can check that the decay process $\phi \rightarrow \gamma \gamma$ has a lifetime that is several orders of magnitude larger than the age of the universe.  From \cite{Cosme:2018nly}, the decay rate of a virtual Higgs to two photons is given by,
\begin{align}
\Gamma_{H^* \rightarrow \gamma \gamma} = \frac{G_F \alpha_{QED}^2 m^3}{128 \sqrt{2} \pi^3} F^2,
\end{align}
where $F \simeq 11/3$ includes all loop contributions from charged fermions and the $W$ bosons.  The Fermi constant $G_F$ will shift due the the shift in the Higgs VEV,
\begin{align}\label{eq:Fermi_shift}
G_F = \frac{1}{\sqrt{2}v^2} = \frac{\bar{G}_F}{1-\beta v_\phi^2/m_h^2}
\end{align}
where $\bar{G}_F \equiv 1/(\sqrt{2}v_\text{ew})$ is the Fermi constant for $\beta = 0$.  Due to the interactions between $\phi$ and the Higgs, there is some mixing, $\epsilon$, given by Eq. (\ref{eq:higgsmixing}) that will suppress the decay rate of $\phi \rightarrow \gamma \gamma$
\begin{align}
\Gamma_{\phi \rightarrow \gamma \gamma} = \epsilon^2 \, \Gamma_{H^* \rightarrow \gamma \gamma}
\end{align}
Taking $\beta v_\phi^2 \ll m_h^2$, the lifetime for $\phi$ is then,
\begin{equation} \label{eq: life_quad}
\tau_{\phi} \sim 10^{43} \, \text{yr} \, \left(\frac{10^{-2}}{\beta}\right)^2 \left(\frac{10^{-6}\,\text{eV}}{m}\right)^3 \left(\frac{10^{4} \, \text{eV}}{v_\phi} \right)^2.
\end{equation}
If we assume that all of the DM in the galaxy is in the form of boson stars, the nonzero expectation value of $\phi$ can be taken to be given by Eq. (\ref{eq:vphi}) inside the boson star and to be equal to zero outside the boson star.  In this case, it can be shown that for $\beta \lesssim 10^{-2}$ and a given $v_\phi$ that can create a recordable clock shift, this lifetime is much larger than the age of the universe.


\end{document}